\documentclass{jfm}
\usepackage{graphicx}
\usepackage{epstopdf, epsfig}
\usepackage{natbib}
\usepackage{amsmath}
\usepackage{mathtools}
\usepackage{subcaption} % SCOTT ADDED THIS
\usepackage{cancel} % Tachado
\usepackage{hyperref}
\usepackage{comment}
\usepackage{tikz}
\usepackage{makecell}
\usepackage{verbatim}
\usepackage{newtxtext}
\usepackage{newtxmath}

\hypersetup{
    colorlinks=true,
    linkcolor=blue,
    citecolor=blue
    }
\usepackage{xcolor}
\usepackage{float}

\shorttitle{Quantifying equilibrium pressure-gradient turbulent boundary layers}
\shortauthor{W.-T. Bi, J. Chen and Z.-S. She}

\title{Quantifying equilibrium pressure-gradient turbulent boundary layers via a symmetry approach}

\author{Wei-Tao Bi\aff{1},
        Jun Chen\aff{1}
 \and Zhen-Su She\aff{1,}
 \corresp{\email{she@pku.edu.cn}}}

\affiliation{\aff{1}State Key Laboratory for Turbulence and Complex Systems and Department of Mechanics and Engineering Science, College of Engineering, Peking University, Beijing 100871, China}

\begin{document}

\maketitle

\begin{abstract}
We propose a theory for predicting the mean velocity and Reynolds shear and normal stresses profiles in the wake region of equilibrium adverse pressure-gradient (PG, APG) turbulent boundary layers (TBLs). The theory is based on the symmetry approach of the structural ensemble dynamics theory of She et al. (J. Fluid Mech., 2017, vol. 827, pp. 322-35; Part 2, 2018, vol. 850, pp. 401-438). Firstly, we explore the PG-induced dilation-symmetry-breaking of the total stress $\tau^+$ to construct a modified defect power law for $\tau^+$ of the equilibrium PG TBLs. Crucially, a PG stress $P_0^+$ is identified, which quantifies the APG-induced total-stress overshoot and is proportional to the Clauser PG parameter $\beta$. The wall-normal location with peak stress is predicted to be $y_{\rm max}^*=y_{\rm max}/\delta=0.455[3\beta/(a+3\beta)]^{2/3}$. With a transformation of $\tau^\#=\tau^+/\tau_{\rm max}^+$, $y^\#=1-0.545(1-y^*)/(1-y_{\rm max}^*)$, the total stress profiles with arbitrary $\beta$ are transformed into an invariant profile of $\tau^{\#}=2[{1+{{({0.49/y^\#})}^4}}]^{-1/4}-2y^{\#3/2}$, which is the ultimate state of the total stress at infinite $\beta$. This transformation is equivalent to the outer scaling of the Reynolds shear stress recently-proposed by Wei \& Knopp (J. Fluid Mech., 2023, vol. 958, A9), but the present one is theoretical, instead of fitting. The Reynolds normal stresses are predicted accordingly based on the similarity of the Reynolds shear and normal stresses in the wake region. Secondly, a defect power law is proposed for the stress and kinetic energy lengths in the wake region. Two critical parameters in the defect power law are identified to depend on $\beta$ and determine the length profiles. One is the exponent $n$, which quantifies the extent of the wake region. The other is $\lambda_{ij}$, which quantifies the relative eddy length in the wake region. With the total stress model and the defect power law of the stress length, the streamwise mean-velocity-deficit profile is predicted. Especially, an invariant mean velocity profile is derived, which describes the ultimate state of the mean velocity in the wake region at infinite $\beta$. This invariant profile is also equivalent to the outer scaling of \cite{wei_knopp_2023}, but the present one is theoretical. The current theory also predicts the variation of the Coles' wake parameter $\Pi$ with $\beta$, in close agreement with the previous empirical relation that correlates hundreds of experimental data. The above predictions are validated with five published DNS, LES, and experimental databases on the equilibrium APG TBLs, and are in accurate agreement with the data over wide ranges of $\beta$ and Reynolds number.
\end{abstract}


\section{Introduction}\label{sec:Introduction}
Turbulent boundary layers (TBLs) on the surfaces of flying, ground, and underwater vehicles, wind and turbo-engine blades, etc., are generally subjected to non-zero pressure gradient (PG). Compared with the canonical zero-pressure-gradient (ZPG) TBL, TBLs with adverse and favorable pressure gradients (APGs/FPGs) are more intriguing and less understood, because of the non-locality nature of pressure, the expanded parametric space (e.g. surface curvature), and the relevant complex flow phenomena such as separation. Owing to their upmost importance in fluid engineering, PG TBLs have received tremendous studies during the past decades. Whereas abundant of knowledge about the structure, mechanism, qualitative and quantitative behaviors of PG TBLs have been concluded, a theoretical description of the mean velocity and turbulent intensity profiles is still to be developed, even for so-called equilibrium TBLs. A consequence is that, most turbulence models, which usually base on calibrations with the law of the wall of ZPG TBL, encounter difficulties in predicting PG TBLs, especially separated TBLs. 

A typical feature of PG TBLs is the classical law of the wall and defect law cease to be valid. To retrieve a similar solution along the streamwise direction, people have studied the equilibrium TBLs \citep{clauser1954turbulent,bradshaw1967turbulence,castillo2001similarity}, which, with constant dimensionless PG, possess a similar velocity-deficit profile in the wake region. This dimensionless PG is the Clauser PG parameter $\beta\equiv(\delta_1/\tau_w)(\partial p /\partial x)|_e$, where $\delta_1$ denotes the displacement thickness of the boundary layer, $\tau_w$ is the mean wall shear stress, $\partial p /\partial x$ is the mean steamwise PG, and subscript $e$ denotes variables at the boundary layer edge. Equilibrium TBLs occur when the free-stream velocity $u_e$ exhibits a power-law variation with the streamwise coordinate $x$ \citep{townsend1961equilibrium,devenport2022equilibrium}, i.e. $u_e\propto |x-x_0|^m$, where $x_0$ is a virtual origin ahead of the self-similar regime, and $m$ is related to $\beta$ by $m=-\beta/(1+3\beta)$ (\cite{skote1998direct} argued that $m=-\beta/[H(1+\beta)+2\beta]$ at finite $Re$s, where $H$ is the shape factor.). Two cases are thus special. $m=0$ (i.e. $\beta=0$) denotes the canonical ZPG TBL with the most thorough investigations \citep{Smits2011annualreview}, and $m=-1$ (i.e. $\beta=-0.5$) denotes the plane sink flow that uniquely possesses similarity over the entire boundary layer \citep{Spalart1987,Jones2001,chenx2017sink,patwardhan2021sink}. For the studies of the PG-effects, attentions belong to equilibrium TBLs with general $m$ (or $\beta$).

The outer-layer similarity of the streamwise mean velocity has commonly been expressed as $u_e^+-u^+=g(y/\delta)$, which is called the K\'{a}rm\'{a}n-Clauser defect law, where $y$ is the wall-normal coordinate, $\delta$ is the boundary layer thickness, superscript plus denotes normalization with the wall units, and function $g$ is PG-dependent and to be determined. However, numerous studies have attempted to identify alternative velocity and length scales. To name a few, \cite{clauser1954turbulent} recognized the problem of scaling on the poorly-defined boundary layer thickness, and instead suggested adopting the integral scale $\Delta\equiv\int_{0}^{\infty}(u_e^+-u^+)dy$, which is known as the Clauser-Rotta boundary layer thickness. \cite{zagarola1998mean} proposed to scale the velocity with the difference between the edge value and the average across the layer, i.e. $(u_e-u)/(u_e-\int_{0}^{\delta}udy/\delta)=g(y/\delta)$, which is referred to as the Zagarola-Smits scaling. \cite{castillo2001similarity} extended the concept of equilibrium TBL and introduced a PG parameter to characterize the velocity profile in the outer flow. They proposed that the outer scales for the velocity deficit, Reynolds normal and shear stresses are, respectively, $u_e$, $u_e^2$, and $u_e^2(d\delta/dx)$. Via a scaling patch approach, \cite{wei_knopp_2023} recently proposed an outer scaling for the streamwise and wall-normal mean velocities, as well as the Reynolds shear stress, of 2D APG TBLs. They constructed a mixing-layer-like scaling for the outer region of the boundary layer. Specifically, they used the maximum Reynolds shear stress location $y_{\rm{max}}$ as the scaling center, $\delta-y_{\rm{max}}$ as the shear-layer width, $u_\delta-u_m$ ($u_m$ is the mean velocity at $y_{\rm{max}}$) as the velocity scale for collapsing the streamwise mean velocity deficit, and the maximum Reynolds shear stress as the characteristic stress for collapsing the Reynolds shear stress. By fitting several datasets of APG TBLs, universal profiles were obtained for the mean velocity and Reynolds shear stress, which collapse well the experimental and numerical data of APG TBLs. The above studies have, either empirically or via a similarity analysis, proposed general similar forms for the velocity and Reynolds stresses profiles in the outer flow. Nevertheless, there seldom have been analytical expressions to describe the profiles. 

Analytical expressions can be derived basically only on the viscous sublayer (which is trivial) and overlap layer. Based on a PG-based velocity scale, \cite{stratford1959prediction} obtained a square-root law for the velocity profile with zero wall shear. \cite{townsend1961equilibrium} extended the theory to PG TBLs with positive wall shear and derived an expression for the mean velocity profile that includes both the square-root and logarithmic parts. \cite{skote2002direct} revisited Townsend's work through an argument of overlap of the inner and outer flows. \cite{ye2017near} further extended the work to compressible flows, specifically oblique shock wave/turbulent boundary layer interactions. Formulations on the wake region are empirical. The most influential work is the law of the wake by \cite{coles1956law}, who conjectured a universal wake function and a PG-dependent wake factor $\Pi$ for describing the outer-layer velocity profiles of quasi-equilibrium PG TBLs. \cite{perry2002streamwise} studied the variation of $\Pi$ in general non-equilibrium TBLs. Although Coles' law is quite successful, there have been debates on the existence and invariance of the logarithmic law embedded in it \citep{Nagib2008,Monty2011,Romero2022jfm}, especially for large $\beta$ at finite $Re$. Besides, there are alternative options for the outer-layer velocity profile, such as the square-root law \citep{stratford1959prediction,Knopp2021jfm} and a mixing-layer-like profile \citep{bradshaw1967turbulence,balantrapu2021}.

To study the equilibrium PG TBLs, plenty of experimental and numerical work have been conducted. As concluded by \cite{devenport2022equilibrium}, the datasets include the experiments of \cite{clauser1954turbulent}, \cite{bradshaw1967turbulence}, \cite{Skare1994exp}, \cite{Jones2001}, and \cite{Sanmiguel2020exp}, as well as the direct numerical simulations (DNSs) and large eddy simulations (LESs) of \cite{Spalart1987}, \cite{kitsios2017direct}, \cite{Bobke2017les}, \cite{Lee2009}, and \cite{Lee2017direct}, to name a few. Although data scatterings occur among different studies, which may be due to sensitivity to initial conditions and the way the PG is imposed, a fundamental view of the effects of APG and FPG on the equilibrium TBLs is now available, which includes the qualitative and quantitative behaviors of the mean-flow properties as well as the local dominant flow structures \citep{devenport2022equilibrium}.

For the mean velocity profiles of APG TBLs, the most prominent feature is the wake region grows in strength and extent, and the $u$ profile develops an inflected form compared to the ZPG case. It also seems established that APG results in a thinner sublayer in terms of wall units, and thus a lowering of the semi-logarithmic region below its expected ZPG level \citep{Monty2011,devenport2022equilibrium}. The APG reduces turbulence levels near the wall as it reduces the mean velocity. However, with wall-unit normalization, ${\left<u'u'\right>}^+$ (where $\left<\cdot\right>$ denotes Reynolds averaging and superscript prime denotes the fluctuation part) near the wall increases significantly with increasing APG. The wall-normal location of the near-wall peak of ${\left<u'u'\right>}^+$ seems to remain at $y^+=15$ \citep{Bobke2017les,devenport2022equilibrium}. In the outer region, turbulence levels increase dramatically with increasing APG, amplifying the outer peak in spectrograms of $u$ and leading to a distinct outer peak in the ${\left<u'u'\right>}^+$ profile, centered near the inflection in the mean velocity profile and mirrored in the other Reynolds stresses \citep{devenport2022equilibrium}. This outer peak is similar to the second peak of ${\left<u'u'\right>}^+$ in ZPG TBLs at high $Re$s \citep{Marusic2010,Sanmiguel2020PRF}. However, the amplification of the large-scale energy by the APG is much more severe than that due to high $Re$ \citep{Monty2011}. Since at least the 1967 work of \cite{bradshaw1967turbulence}, the eddies associated with these outer region motions have been hypothesized as being similar to those of a free mixing layer. Budgets of the turbulent kinetic energy reveal the mechanisms responsible for the outer-layer profiles of $u$ and Reynolds stresses \citep{Skare1994exp,Bobke2017les,Gungor2022}. In the APG case the production peaks both very close to the wall and in the outer region. Diffusion transports the turbulence kinetic energy away from the outer peak and into the regions close to the wall and the edge of the boundary layer. Convection is substantial in the outer half of the boundary layer because of the increase in turbulence kinetic energy along streamlines being entrained into the boundary layer and drawn into the turbulence producing outer region \citep{devenport2022equilibrium}.

In terms of the turbulence structure, as reviewed by \cite{devenport2022equilibrium}, numerical simulations indicate that equilibrium APG reduces the number and increases the spacing of low-speed streaks in the sublayer. Correlation analysis suggests that the streamwise length scales near the wall increase under mild APGs but reduce when the APGs are stronger. The angles (relative to the wall) of burst and sweep events increase with increasing APG. Whereas the frequency of bursts in the inner region decreases considerably in APG TBL, that of sweeps increases compared to ZPG TBL. Further away from the wall, the trains of hairpin vortices and the associated elongated low-momentum regions are more prominent in the outer region of the APG TBL than those of the ZPG case \citep{skote1998direct,Lee2009}. The angle of these structures also significantly increases with increasing APG. In the outer region of APG TBL, the frequency of second quadrant (ejection type) shear-stress generating motions is much higher, and the ratio of the streamwise to spanwise length scales decreases remarkably. At the top end of the boundary layer there are very-large-scale-motions (termed superstructures) that extend over distances about ten times of $\delta$, composed of related groups of the smaller structures \citep{Smits2020}. These outer-layer structures are responsible for the outer peak in the ${\left<u'u'\right>}^+$ profile.

With the above fruitful knowledge, a theoretical description of the whole profiles of both the mean velocity and Reynolds stresses is probable for the equilibrium PG TBLs, which, however, requires a new framework beyond the conventional local perturbation analysis and empirical constructions. The present work intends to propose such a description based on the symmetry approach of the structure ensemble dynamics (SED) theory.

Recently, \cite{she2010new,she2017quantifying} proposed the SED theory for the canonical wall-bounded turbulent flows including pipe, channel, and planar ZPG TBL. In the SED, the constraint imposed by the wall on the turbulence is expressed through one stress and three kinetic energy length functions that exhibit multilayered dilation symmetries with respect to the wall distance, forming so-called multilayer structures (MLSs) that arise due to the conversion of balance mechanisms away from the wall in the transport equations of turbulent fluctuations \citep{chen2018quantifying}. By introducing a simple dilation-symmetry-breaking ansatz to represent the transition of dilation symmetries between adjacent layers, the SED provides excellent analytical descriptions of experimentally and numerically observed mean velocity and Reynolds stress components profiles of canonical wall turbulence. Furthermore, for the non-canonical TBLs with various effects such as compressibility, roughness, and PG, the SED assumes that the dilation symmetry is preserved because of the dominant role of the wall constraint, but the MLS is limitedly perturbed by the effects. Thus, with variable MLS parameters to quantify the perturbed MLS, the SED solution can be extended to describe non-canonical TBLs, such as rough pipe TBLs \citep{she2012multi}, transitional boundary layers \citep{xiao2019symmetry}, and atmospheric TBLs \citep{ji2021analytic}. In this paper, the symmetry approach of the SED is extended to describe the equilibrium PG TBLs. 

Here, a modified defect law is constructed for describing the PG effects on the total stress profiles of the equilibrium PG TBLs. The PG effects on the stress and kinetic energy lengths in the wake region are described with a defect power law characterized by two critical parameters. Consequently, with the symmetry-based formulations of the total stress and length functions, the mean velocity and Reynolds shear and normal stresses are predicted and validated with the recent DNS and experimental data. 

The paper is organized as follows. Section \ref{sec:theory} presents the details of the current theory with a brief of review of the conventional analysis. Section \ref{sec:validation} validates the descriptions of the theory, in the order of the total stress, Reynolds shear and normal stresses, and streamwise mean velocity. Section \ref{sec:conclusion} concludes the study. 

\section{Theory}\label{sec:theory}
\subsection{Balance equation and closure strategy}\label{subsec:closure_strategy}
The streamwise mean momentum equation of a two-dimensional (2D) incompressible planar TBL reads
 \begin{equation}
  u\frac{\partial u}{\partial x}+v\frac{\partial u}{\partial y}  = -\frac{1}{\rho} \frac{\partial p}{\partial x}+\nu \left(\frac{\partial^2 u}{\partial x^2}+\frac{\partial^2 u}{\partial y^2}\right)-\frac{\partial {\left<u'u'\right>}}{\partial x}-\frac{\partial {\left<u'v'\right>}}{\partial y},
  \label{eq:x_momentum}
\end{equation}
where $x$ and $y$ denote the streamwise and wall-normal coordinates, $u$ and $v$ represent the streamwise and wall-normal mean velocity components, respectively, with $u'$ and $v'$ representing the corresponding velocity fluctuations, $\rho$ is density, $p$ is mean static pressure, $\nu$ is kinematic molecular viscosity, and $\left<\cdot\right>$ denotes Reynolds averaging. The wall-normal integration of (\ref{eq:x_momentum}) gives
 \begin{equation}
  \tau\equiv\nu \frac{\partial u}{\partial y}-{\left<u'v'\right>}=\nu \frac{\partial u}{\partial y}\Big|_w +\int_{0}^{y}{\left(u\frac{\partial u}{\partial x}+v\frac{\partial u}{\partial y}+\frac{1}{\rho} \frac{\partial p}{\partial x}-\nu \frac{\partial^2 u}{\partial x^2}+\frac{\partial {\left<u'u'\right>}}{\partial x}\right)dy},
  \label{eq:tau}
\end{equation}
where subscript $w$ denotes the variables on the wall, and $\tau$ is the total stress. In the integral on the right hand side of (\ref{eq:tau}), the convection (the first and second terms) and streamwise PG are the major components contributing to the $\tau$ profile, except in the vicinity of incipient separation where the streamwise gradient of $\left<u'u'\right>$ plays a finite role.

A conventional strategy to predict the mean velocity profile through (\ref{eq:tau}) models the $\tau$ and $-\left<u'v'\right>$ profiles in the viscous sublayer and overlap layer, respectively. In the viscous sublayer, the convection is negligible and $\tau$ has the approximation of
\begin{equation}
  \tau^+=1 +P_w^+ y^+,
  \label{eq:tau_nearwall}
\end{equation}
where superscript plus denotes normalization with the wall units, $P_w^+ = [\nu/(\rho u_\tau^3)]({\partial p}/{\partial x})_w$ ($u_\tau$ is the friction velocity) is the PG parameter (note that $P_w^+$ is related to $\beta$ via $P_w^+=\beta/{\delta_1}^+$). Because $-\left<u'v'\right>$ is also negligible, the $u^+$ profile in the viscous sublayer can be derived from (\ref{eq:tau}) and (\ref{eq:tau_nearwall}) as below 
\begin{equation}
  u^+=y^+ +P_w^+ {y^+}^2/2,
  \label{eq:U_nearwall}
\end{equation}
which adapts to general PG TBLs. Within the overlap region, the Reynolds shear stress dominates the total stress and is conventionally modeled with the mixing length $\ell_{12}$: 
\begin{equation}
  -{\left<u'v'\right>}^+={\ell_{12}^+}^2{S^+}^2,
  \label{eq:mixinglengthhypo}
\end{equation} 
where $S^+=\partial{u^+}/\partial{y^+}$ is the mean shear. According to Prandtl's mixing length hypothesis \citep{prandtl1925ausgebildete}, $\ell_{12}^+=\kappa_{12} y^+$. In the canonical ZPG TBL, $\kappa_{12}=\kappa$ with $\kappa$ being the von K\'{a}rm\'{a}n constant. By further assuming $\tau^+$ to preserve a linear profile, i.e. $\tau^+=1 +\lambda y^+$ where $\lambda$ is an effective PG parameter (note that $\lambda$ is close to but not necessarily equal to $P_w^+$.), $u$ in the overlap region can be solved as \citep{townsend1961equilibrium,knopp2015investigation}
\begin{equation}
  u^+=\frac{1}{\kappa_{12}}\left[{\rm ln} y^+-2{\rm ln}\frac{\sqrt{1+\lambda {y^+}}+1}{2}+2(\sqrt{1+\lambda {y^+}}-1)\right]+B,
  \label{eq:U_overlap}
\end{equation}
where $\kappa_{12}$, $\lambda$, and $B$ are PG-dependent parameters. In the case of ZPG, (\ref{eq:U_overlap}) is reduced to the celebrated logarithmic law of the wall. Therefore, (\ref{eq:U_overlap}) is called ``generalized overlap law''. Note that (\ref{eq:U_overlap}) can also be derived via a similarity argument \citep{skote2002direct} traced to \cite{Millikan1938}, which is thought to be more insightful sometimes. For this reason, (\ref{eq:U_overlap}) should be appropriate for equilibrium PG TBLs.

For a long time, the other parts of the boundary layer are lack of theoretical description. The most well-known formulation is about the outer-layer profile of $u^+$, proposed empirically by \cite{coles1956law} and called Coles' law of the wake, which reads
\begin{equation}
  u^+=\frac{1}{\kappa}{\rm ln}y^++B+\frac{2{\rm \Pi}}{\kappa}f\left(y^*\right),
  \label{eq:U_wake}
\end{equation}
where $\kappa=0.41$, $B=5.0$, $y^*=y/\delta$, $\delta$ is the boundary layer thickness, the wake function $f$ has a popular form of $f(\eta)={\rm sin}^2\left(\pi\eta/2\right)$, and Coles' wake parameter ${\rm \Pi}$ depends on $\beta$. (\ref{eq:U_wake}) was argued to apply to near equilibrium PG TBLs.

Note that (\ref{eq:tau_nearwall}), (\ref{eq:U_nearwall}), (\ref{eq:U_overlap}), and (\ref{eq:U_wake}) are not invariant but dependent on the PG. Therefore, there has no PG-similarity in the above formulations of the PG TBL, which is different from the $Re$-similarity of the ZPG TBL.

Our study follows the conventional total-stress mixing-length closure stategy. Nevertheless, we formulate the $\tau^+$ and $\ell_{12}^+$ profiles of the equilibrium PG TBLs based on the symmetry approach of the SED theory. To be concise, we focus only on the wake region at the moment, aiming to describing the similarity of the flows, especially describing the intriguing inflected velocity-deficit profile as well as the profiles of the Reynolds stresses with distinct outer peaks. Formulations for the inner region can be conducted similarly, but will be reported elsewhere.   

With $\tau^+$ and $\ell_{12}^+$, the mean shear $S^+$ can be solved through (\ref{eq:tau}) and (\ref{eq:mixinglengthhypo}) as
 \begin{equation}
  S^+ = \frac{{ - 1 + \sqrt {4{\tau ^{\rm{ + }}}{{ {\ell_{12}^+}}^2}+ 1} }}{{2{{ {\ell_{12}^+}}^2}}}\approx\frac{\sqrt{\tau^+}}{\ell_{12}^+}, \quad\quad y^+\gg 0.
  \label{eq:mean_shear}
\end{equation}
The $u^+$ profile in the defect form can be integrated to be
 \begin{equation}
  u_\delta^+-u^+=  \int_{y^*}^{1} {S^+\delta^+ \mathrm{d} \left(\frac{y^+}{\delta^+}\right)},
 \label{eq:udefect_model}
 \end{equation}
where $u_\delta^+$ denotes the mean velocity at the boundary layer thickness. 

The Reynolds shear stress profile can be predicted by (\ref{eq:mixinglengthhypo}). Furthermore, the Reynolds normal stresses can similarly be predicted with the so-called kinetic energy lengths (which are introduced by the SED as analogies of the mixing length) as below:
\refstepcounter{equation}
$$
  {\left<u'u'\right>}^+={\ell_{11}^+}^2 {S^+}^2,\quad
  {\left<v'v'\right>}^+={\ell_{22}^+}^2 {S^+}^2,\quad
  {\left<w'w'\right>}^+={\ell_{33}^+}^2 {S^+}^2,
  \eqno{(\theequation{\mathit{a},\mathit{b},\mathit{c}})}\label{eq:kinetic_energy}
$$
\noindent where $w$ denotes the spanwise velocity, and $\ell_{11}^+$, $\ell_{22}^+$, and $\ell_{33}^+$ represent the streamwise, wall-normal, and spanwise kinetic energy lengths, respectively. Different from the Prandtl's mixing length argument, these lengths with together $\ell_{12}^+$ are interpreted as the characteristic lengths of eddies responsible for transporting the corresponding turbulent quantities. Because of this interpretation and also because of the extension to the Reynolds normal stresses, $\ell_{12}^+$ is renamed ``stress length'' in the SED and hereinafter in this article.

\subsection {A modified defect power law of the total stress}\label{subsec:tau_model}
Here we present a novel formulation for the $\tau^+$ profile of the equilibrium PG TBL. Because the similarity analysis of the balance equations has been well documented in quite a few studies, such as \cite{Townsend1976,castillo2001similarity,maciel2006self,devenport2022equilibrium}, to name a few, we do not start from the very beginning of investigating the similarity forms of $\tau^+$ and the length functions. Instead, we directly construct the $\tau^+$ and length models that meet the requirements of similarity. Different from \cite{knopp2015investigation}, we also do not investigate the detailed items contributing to $\tau^+$, but seek the model by exploring the dilation symmetry of $\tau^+$. 

For the canonical wall turbulence, $\tau^+$ possesses the following defect power law:
\begin{equation}
\tau^+= 1-{y^*}^{\alpha},\quad \quad y\leq\delta,
  \label{eq:tau_ZPG}
\end{equation}
where for channel flows $\delta$ represents the half-width of the channel, and for pipe flows $\delta$ represents the radius of the pipe, and $\alpha$ is a constant. $\alpha$ is exactly unity in channel and pipe flows because of the zero convection. In TBLs, \cite{Degraaff2000JFM} have shown that $\alpha$ deviates from unity. By inspecting experimental and numerical data, \cite{chen2016analytic} have proposed that $\alpha=3/2$ for the canonical ZPG TBLs. Note that (\ref{eq:tau_ZPG}) is invariant in the outer spatial coordidate of $y^*$, independently of $x$, thus obeying the similarity requirement. 

In the case of PG TBLs, (\ref{eq:tau_ZPG}) is no longer applicable and must be modified to account for the PG effect. To extend (\ref{eq:tau_ZPG}) to equilibrium PG TBLs, we make the following two assumptions. The first assumption is that the PG does not modify the decaying behavior of $\tau^+$ when $y$ approaches the potential flow. In other words, $\tau^+$ near the boundary layer edge preserves the defect power law, i.e.
 \begin{equation}
  \tau ^ +  = \left(1+P_0^+\right) \left[1-{y^*}^{3/2}\right],\quad \quad 0\ll y\leq\delta,
  \label{eq:tau_bulk}
\end{equation}
where $P_0^+$ is a PG-induced characteristic stress (called PG stress) that represents a measure of the Reynolds stress overshoot owing to APG. In the case of ZPG, (\ref{eq:tau_bulk}) reverts to (\ref{eq:tau_ZPG}). At the moment we have not derived an explanation of the physical basis of the assumption. Therefore, (\ref{eq:tau_bulk}) can be understood as a simple extension of (\ref{eq:tau_ZPG}). However, as shown by \cite{DuTT2023arxiv}, the assumption is invalidated in reattached PG TBLs in which there are intense turbulent motions convected from upstream separated shear layer. Consequently, further studies should be performed to elucidate the physical basis as well as the application domain of the assumption.   

The second assumption is about the structure of the $\tau^+$ profile, which, in equilibrium PG TBLs, is assumed to be a connection of the near-wall linear profile of $\tau^+=1+\lambda y^+$ to the near-edge profile (\ref{eq:tau_bulk}). According to the SED, these two profiles are formed owing to a dilation-symmetry-breaking induced by a non-zero PG. Therefore, with a universal ansatz proposed by the SED for describing the dilation-symmetry-breaking process, a unified profile can be constructed for $\tau^+$ of equilibirum PG TBLs, as shown below
\begin{equation}
\left. \begin{array}{ll}
\displaystyle \tau^+=(1+P_0^+)\left[1-{y^*}^{3/2}\right]+P_0^+\left\{{{\left[ {1 + {{\left( {\frac{y_P^+}{y^+}} \right)}^4}} \right]}^{-1/4} - 1} \right\},\quad \quad y\leq\delta,\\[8pt]
\displaystyle  \tau^+=0,\quad \quad y>\delta,
 \end{array}\right\}
  \label{eq:tau_PG}
\end{equation}
where the ansatz reads $[1+(y_P^+/y^+)^4]^{-1/4}$, and $y_P^+=P_0^+/\lambda$ denotes the transition center between the near-wall and near-edge profiles. Indeed, near the wall where $y^+\ll y_P^+$, (\ref{eq:tau_PG}) is reduced to $\tau^+=1+\lambda y^+$, and near the boundary-layer edge where $y^+\gg y_P^+$, (\ref{eq:tau_PG}) is reduced to (\ref{eq:tau_bulk}). Besides, for the canonical ZPG TBLs in which $P_0^+=0$, (\ref{eq:tau_PG}) is reduced to (\ref{eq:tau_ZPG}).

The second assumption requires that (\ref{eq:tau_PG}) applies only to equilibrium PG TBLs. Nonequilibrium PG TBLs are subjected to streamwise varying PG, responding to which are different in speed between the inner and outer flows \citep{devenport2022equilibrium}. Consequently, the transition from the near-wall profile to near-edge profile must be distorted and dependent on the flow history. Especially in the case of a TBL subjected to a sudden and strong APG, an internal boundary layer grows from the wall \citep{Parthasarathy2023}, thus undermining the two-layer structure of $\tau^+$ assumed by the current approach. Further extending (\ref{eq:tau_PG}) to describe nonequilibrium PG TBLs can be conducted by quantifying the history effects, as shown in our previous study on the APG TBL in front of a separation bubble and the reattached TBL behind the bubble \citep{DuTT2023arxiv}. In a companion paper of this study, which will be submitted later, the extension is proposed for several typical nonequilibrium pocesses such that a general non-equilibrium PG TBL can be analyzed piecewise.

(\ref{eq:tau_PG}) should meet the requirements of similarity. (\ref{eq:tau_bulk}) indicates that $P_0^+$ should keep invariant with $x$, although it can depend on $\beta$ and $Re_\tau$ (i.e. the friction Reynolds number). If neglecting the difference between $\lambda$ and $P_w^+$, we have $y_P^+=P_0^+\delta_1^+/\beta$. Because $y_P^+/y^+=y_P^*/y^*$ (where $y_P^*=y_P/\delta$), for (\ref{eq:tau_PG}) to be similar in the outer spatial coordinate of $y^*$, $y_P^*=(P_0^+/\beta)(\delta_1/\delta)$ should be independent of $x$. In other words, $\delta_1$ should be proportional to $\delta$ during the streamwise development, which is a known feature of equilibrim PG TBLs. Therefore, (\ref{eq:tau_PG}) obeys the similarity requirements if $P_0^+$ keeps constant streamwise.

(\ref{eq:tau_PG}) can predict the peak total stress of an equilibrium APG TBL. The magnitude and wall-normal location of the total stress peak, denoted by $\tau_{\rm{max}}^+$ and $y_{\rm{max}}^*$ (where $y_{\rm{max}}^*=y_{\rm{max}}/\delta$), respectively, cannot be solved explicitly from (\ref{eq:tau_PG}). However, a rather accurate estimation (with error less than 1\%) of $y_{\rm{max}}^+$ can be derived as below 
\begin{equation}
  y_{\rm{max}}^* \approx\frac{3}{2+3.57A}y_P^*,
  \label{eq:y_max_tau}
\end{equation} 
where 
\begin{equation}
  A=\frac{1+P_0^+}{P_0^+}{y_P^*}^{3/2}.
  \label{eq:A}
\end{equation} 
(\ref{eq:y_max_tau}) is obtained by assuming $y_{\rm{max}}^*-y_P^*$ is small, such that a first-order Taylor-series approximation can be applied for solving $y_{\rm{max}}^+$ with the equation $d\tau^+/dy^+=0$. Accordingly, $\tau_{\rm{max}}^+$ can be computed through (\ref{eq:tau_PG}) and (\ref{eq:y_max_tau}). The expression is complicated, but one can derive that
\begin{equation}
  \frac{\tau_{\rm{max}}^+-1}{P_0^+}=g(A),
  \label{eq:stress_max}
\end{equation} 
where $g$ is a deterministic function of $A$, and is not shown here for brevity.

For general PG TBLs, the definition of $\delta$ can be problematic \citep{vinuesa2016determining}. Here, $\delta$ in (\ref{eq:tau_PG}) was estimated from the mean velocity profile based on the condition of vanishing mean velocity gradient (with threshold of $S^+=5\times10^{-3}$), as evaluated by \cite{vinuesa2016determining}. Another point is that, although $\lambda\approx/P_w^+$, a better description of the data with limited $Re_\tau$ can be achieved for (\ref{eq:tau_PG}) if $y_P^+$ is set as a free parameter. The reason for this departure is guessed due to a finite Reynolds number effect, as discussed in section \ref{subsec:total_stress_profile}. Therefore, in the validation of (\ref{eq:tau_PG}), both $P_0^+$ and $y_P^+$ are estimated from the $\tau^+$ profile via a least square fit method.

Finally, it is interesting to compare (\ref{eq:tau_PG}) with the model of \cite{Perry1994taumodel}. Assuming the mean velocity profile is given by the Coles' law of the wake, \cite{Perry1994taumodel} derived a model for $\tau$ with a rather complicated analytical expression. They applied the model to TBLs where the streamwise derivative of the Coles' wake parameter is not too large (thus the flows are quasi-equilibrium). Comparing with their model, the current formula is explicit, much simpler, and easy to implement.
  
\subsection {Outer-layer dilation symmetry of the stress and kinetic energy lengths}\label{subsec:length_model}
Now we study the profiles of the stress length and kinetic energy lengths. From the physical viewpoint, APG tends to reduce the mean shear, relax the wall constraint, and elevate the eddies \citep{maciel2017coherent}. Therefore, comparing with the ZPG TBL, the APG TBL possesses a thicker wake region, within which the eddies are unattached to the wall and their characteristic lengths are fractions of $\delta$ and approximately independent of $y$. That is, $\ell_{ij}\approx\lambda_{ij}\delta$ ($i,j=1,2,3$), where $\lambda_{ij}$ are the proportional coefficients quantifying the relative eddy lengths in this constant-length region \citep{East1979}. Beneath the constant-length region, the conventional wisdom models $\ell_{ij}$ with the Prandtl's mixing length hypothesis, i.e. $\ell_{ij}=\kappa_{ij}y$. Now, we refine this two-layer model by invoking a defect power law, as proposed by \cite{she2017quantifying} and shown below,
\begin{equation}
  \ell_{ij}=\lambda_{ij}\delta(1-r^n), \quad i,j=1,2,3,
  \label{eq:length_out}
\end{equation}
where $r=1-y^*$ is the defect spatial coordinate with the origin at the boundary layer thickness and the unity on the wall, and the exponent $n$ is a PG-dependent integer and is assumed to be the same for all the lengths, as proposed by the SED for the canonical flows. With the defect power law, (\ref{eq:length_out}) can describe the entire wake region. Indeed, when $r$ is smaller than one (as shown later, $n$ is from 4 to less than 15 in the APG TBLs studied currently), (\ref{eq:length_out}) reverts to $\ell_{ij}\approx\lambda_{ij}\delta$. When $r$ approaches one, i.e. the overlap layer is reached, $\ell_{ij}\approx n\lambda_{ij}y$, agreeing with the conventional assumption of $\ell_{ij}=\kappa_{ij}y$ with $\kappa_{ij}=n\lambda_{ij}$. The exponent $n$ in (\ref{eq:length_out}) characterizes the relative extent of the wake region: the larger $n$ is, the thicker the wake region, and the stronger the APG. 

To meet the similarity requirements, $n$ and $\lambda_{ij}$ (or $\kappa_{ij}$) in (\ref{eq:length_out}) should be invariant with $x$, although they certainly depend on $\beta$ and may depend on finite $Re_\tau$.

Note that a recent study by \cite{subrahmanyam2022universal} proposed a different construction for $\ell_{12}$. In the outer flow their formulation is reduced to $\ell_{12}=\kappa y\left[1+\left(\frac{y}{b\delta}\right)^m\right]^{-1/m}$, where $\kappa$, $m$, and $b$ are empirical parameters. One finds that, insteading employing a defect power law as in the current approach, they applied the dilation-symmetry-breaking ansatz to describe the transition of $\ell_{12}$ from the Prandtl's linear law in the overlap layer to the constant-stress-length layer in the wake region. The transition center is located at $b\delta$, which defines a boundary between the overlap layer and the wake region. In the constant-stress-length layer, $\ell_{12}\approx b\kappa\delta$. Comparing that with (\ref{eq:length_out}), we can derive $b=1/n$. Therefore, the wake region extends towards the wall when the APG is increased. In the proposal of \cite{subrahmanyam2022universal}, the extent of the wake region is quantified by $b$ and $m$ altogether. The latter represents the transtion sharpness in the ansatz, and should play a limited role in affecting the extent of the wake region.

Another issue that may question the defect power law of (\ref{eq:length_out}) is the hypothesis that the eddies associated with the outer-layer motions are similar to those of a free mixing layer \citep{bradshaw1967turbulence,balantrapu2021,wei_knopp_2023}. As is well-known, in a canonical mixing layer the stress length decreases when $y$ approaches the potential flow. However, to our observation, in the equilibrium APG TBLs the stress length near the boundary layer edge rather keeps constant than displays a clear decreasing trend with increasing $y$, even when the APG is considerably strong. This is in contrast to the reattached TBL, where intense turbulent motions are convected from the upstream separated shear layer, thereby the outer-layer flow indeed is similar to a free mixing layer \citep{DuTT2023arxiv}.   

\section{Validation}\label{sec:validation}
Tremendous amount of experimental and numerical data have been accumulated in the literature for PG TBLs. To assess the current theory we have collected several published databases on equilibrium and near equilibrium PG TBLs, which are listed in table \ref{tab:cases}. Note that, to apply the theory, the $\tau^+$ profile (or equivalently the $u^+$ and ${-\left<u'v'\right>}^+$ profiles together) should be provided in a study, which is because the present theory is based on the wall-unit normalization of the lengths, velocities, and stresses. The databases complying with this requirement are quite a few. As shown in table \ref{tab:cases}, the databases cover wide ranges of $Re_\theta$ ($1250\sim50980$) and $\beta$ ($0\sim21.2$), to the best of the current status of DNS/LES and experiment. Two ZPG TBLs (not listed in table \ref{tab:cases}) are also analyzed. One is the DNS of \cite{Schlatter2009dns}, which possesses matched $Re_\theta$ with those of \cite{Bobke2017les}, and the other is the LES of \cite{Eitel-Amor2014}, which possesses matched $Re_\theta$ with those of \cite{Pozuelo2022les}. 

\begin{table}
  \begin{center}
\def~{\hphantom{0}}
  \begin{tabular}{lcccccc}
      Source & Method  & Type  & $Re_\theta$   &  $\beta$ & $H$ & \makecell{Presented\\quantities}  \\[3pt]
     \cite{Pozuelo2022les} & LES  & near E & 8700 & 1.4 & 1.5 & \makecell{$u^+$,${\left<u'v'\right>}^+$\\${\left<u'u'\right>}^+$,${\left<v'v'\right>}^+$,${\left<w'w'\right>}^+$}\\
     \cite{Lee2017direct} & DNS     & E & \makecell{1320,1605\\2180,2840} & \makecell{0,0.73\\2.2,9.0} & \makecell{1.45,1.56\\1.72,1.98} & \makecell{$u^+$\\${\left<u'v'\right>}^+$,${\left<u'u'\right>}^+$} \\
     \cite{Bobke2017les} & LES  & E & \makecell{2870,3360} & \makecell{1,2} & \makecell{1.56,1.68} & \makecell{$u^+$,${\left<u'v'\right>}^+$\\${\left<u'u'\right>}^+$,${\left<v'v'\right>}^+$,${\left<w'w'\right>}^+$}\\
     \cite{Lee2009} & DNS     & E & \makecell{1410,1250\\1250,1350} & \makecell{0,0.25\\0.73,1.68} & \makecell{1.47,1.55\\1.62,1.84} & \makecell{$u^+$,${\left<u'v'\right>}^+$\\${\left<u'u'\right>}^+$,${\left<v'v'\right>}^+$,${\left<w'w'\right>}^+$} \\
     \cite{Skare1994exp} & Exp   & E  & 50980 & 21.2 & 1.998 & \makecell{$u^+$,${\left<u'v'\right>}^+$\\${\left<u'u'\right>}^+$,${\left<v'v'\right>}^+$,${\left<w'w'\right>}^+$} \\
  \end{tabular}
  \caption{Databases of the equilibrium PG TBLs used for validating the current theory. E in Type denotes equilibrium TBL.}
  \label{tab:cases}
  \end{center}
\end{table}

\subsection{The total stress profile}\label{subsec:total_stress_profile}
\begin{figure}
  \begin{subfigure}[b]{0.5\linewidth}
    \centering
    \includegraphics[width=\linewidth]{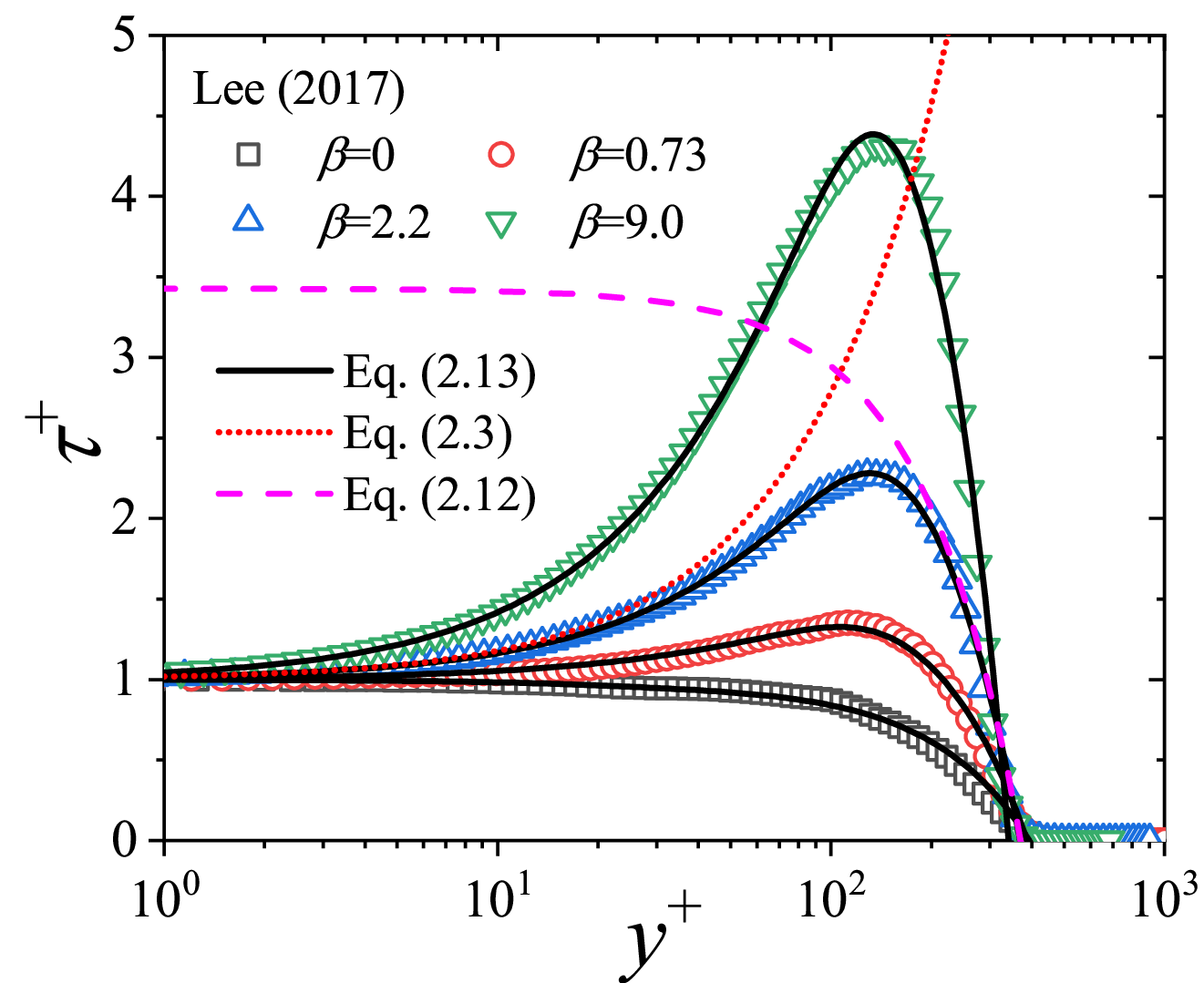}
    \caption{}
  \end{subfigure}
  \begin{subfigure}[b]{0.5\linewidth}
    \centering
    \includegraphics[width=\linewidth]{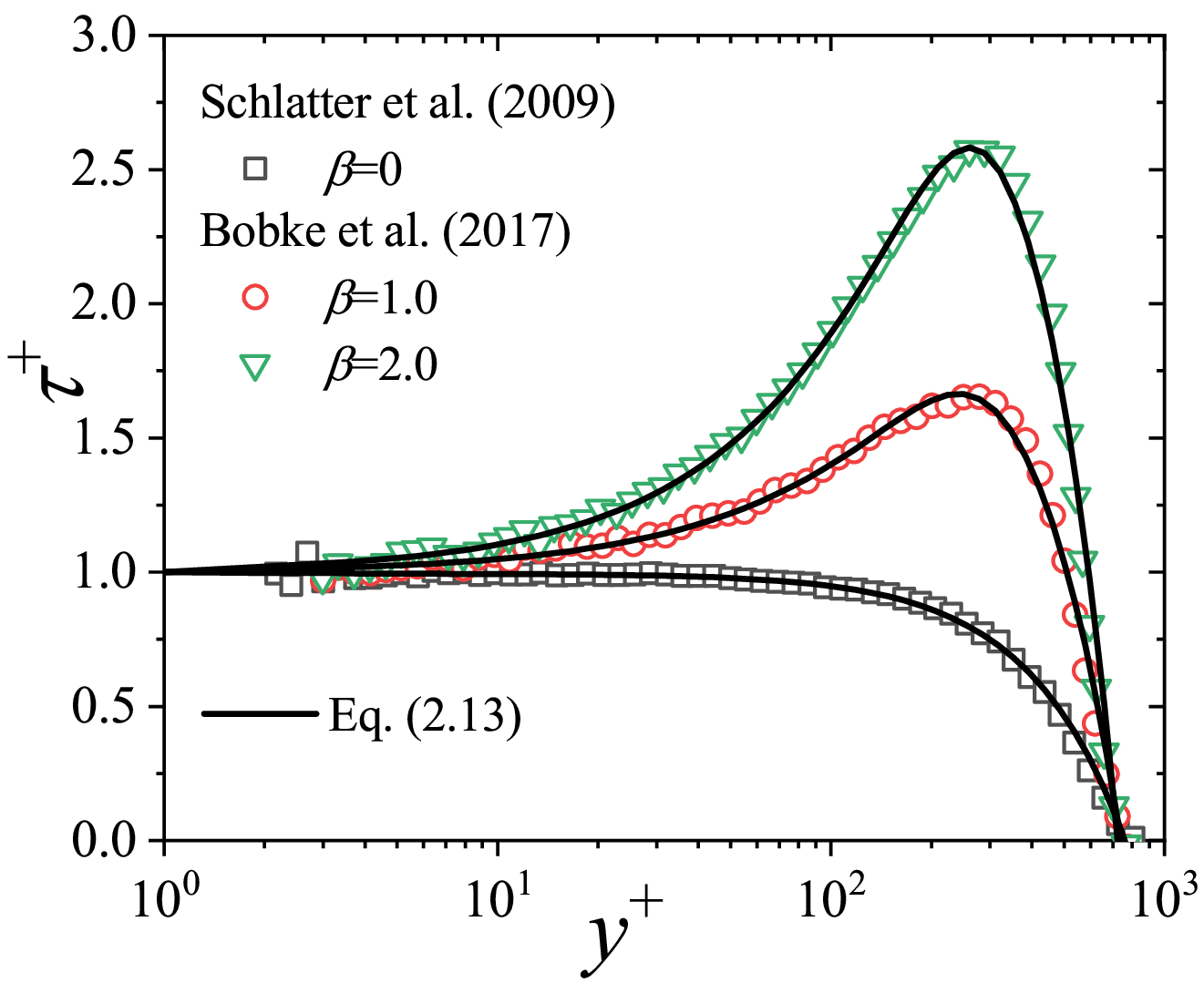}
    \caption{}
  \end{subfigure}  
  \begin{subfigure}[b]{0.5\linewidth}
    \centering
    \includegraphics[width=\linewidth]{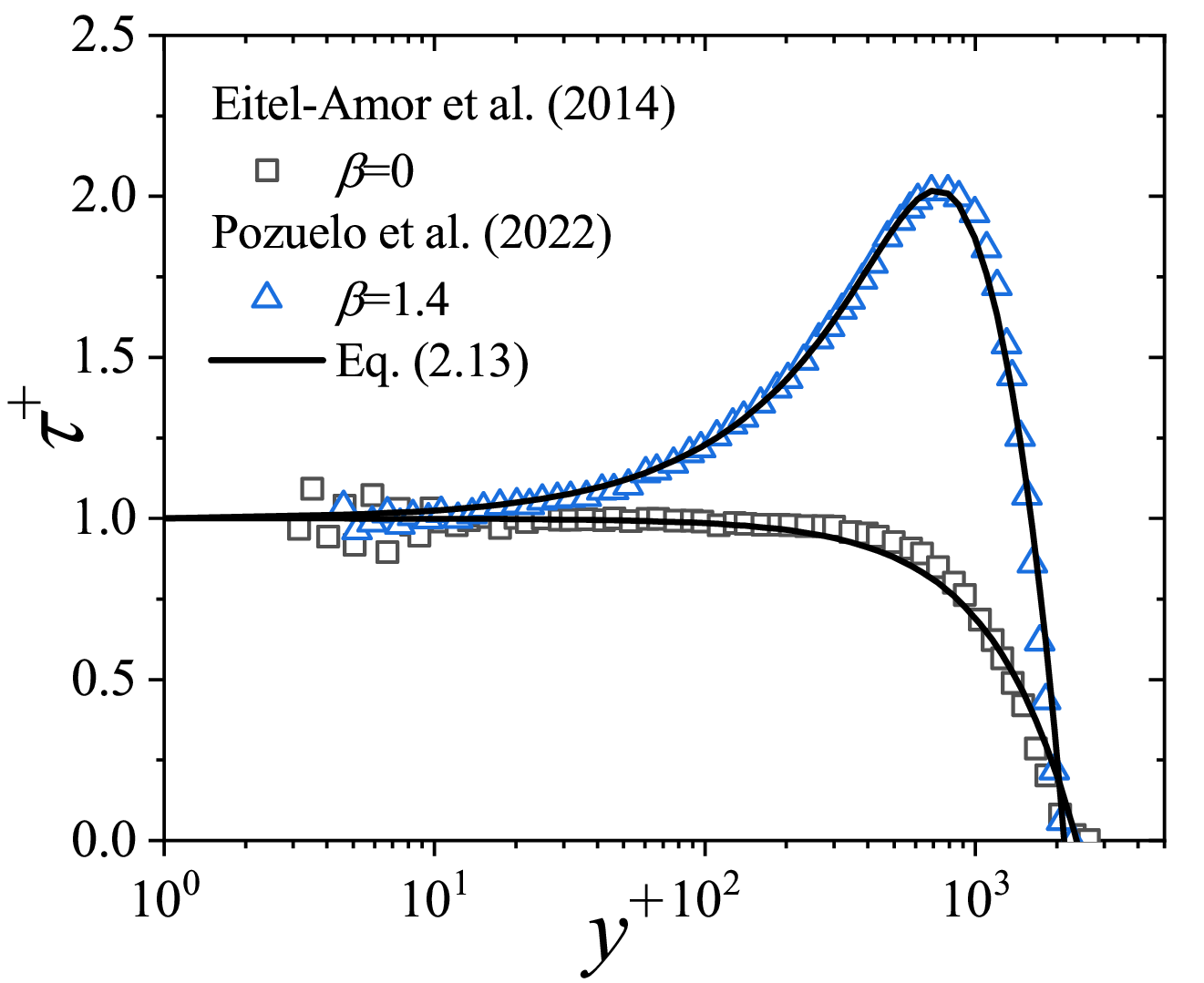}
    \caption{}
  \end{subfigure}  
  \begin{subfigure}[b]{0.5\linewidth}
    \centering
    \includegraphics[width=\linewidth]{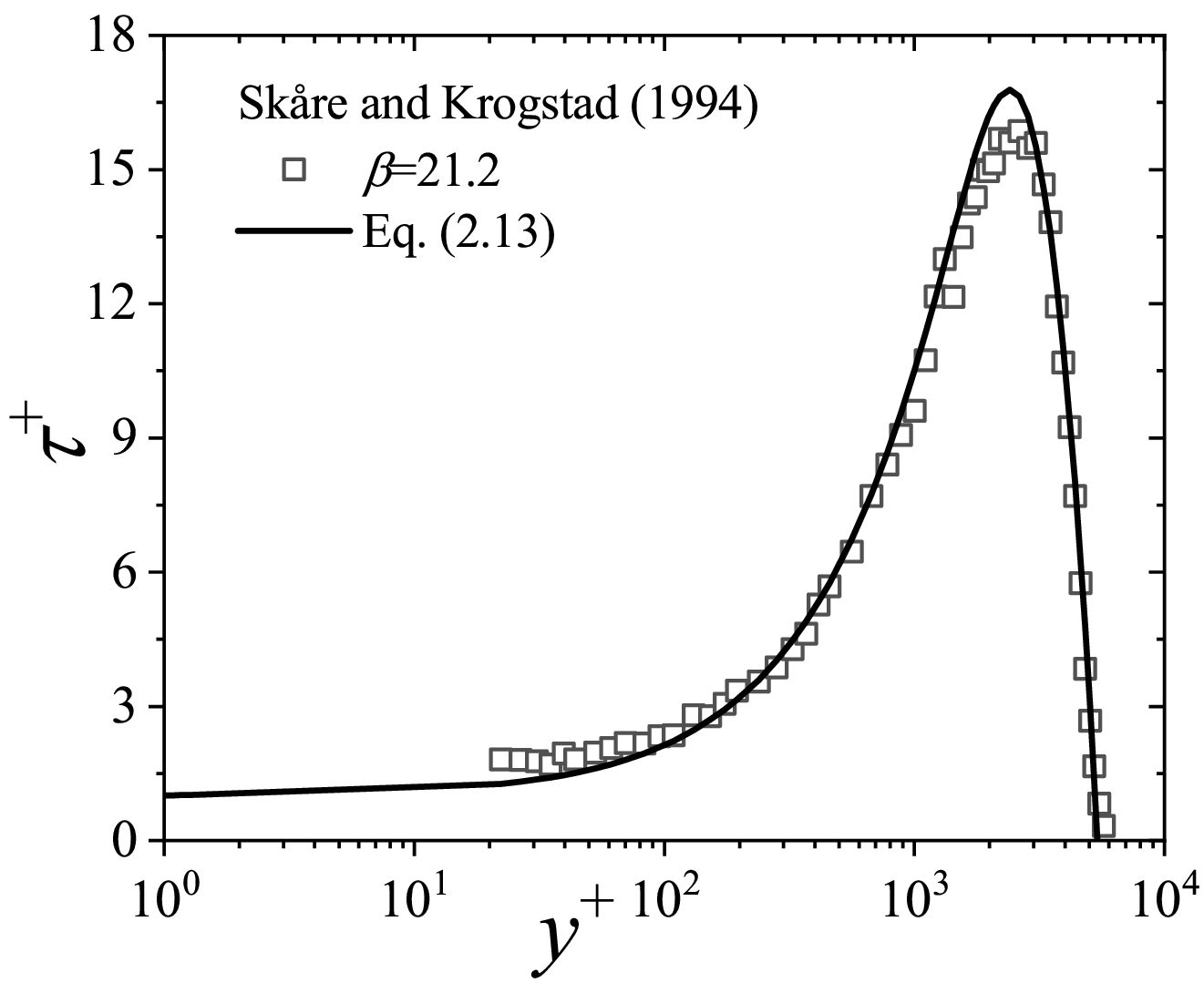}
    \caption{}
  \end{subfigure}  
  \caption{The total stress profiles of the equilibrium APG TBLs in table \ref{tab:cases}, comparing with the current models. The data scatterings in $y^+<10$ in (b) and (c) are due to errors in computing $S^+$ from the mean velocity profiles digitized from the literature.}
  \label{fig:tau_result_E}
\end{figure}

The total stress model (\ref{eq:tau_PG}) is validated in figure \ref{fig:tau_result_E} for the databases in table \ref{tab:cases}. As aforementioned, both $P_0^+$ and $y_P^+$ are estimated via a least square fit of the entire profile of $\tau^+$. Therefore, the validation is posterior. In figure \ref{fig:tau_result_E}(a) the $\beta=2.2$ case of \cite{Lee2017direct} displays the construction of (\ref{eq:tau_PG}). The dilation-symmetry-breaking ansatz makes a smooth transition from the near-wall profile (\ref{eq:tau_nearwall}) to the near-edge profile (\ref{eq:tau_bulk}), such that (\ref{eq:tau_PG}) captures the entire profile of $\tau^+$. As shown in figure \ref{fig:tau_result_E}, (\ref{eq:tau_PG}) rather accurately describes the $\tau^+$ profiles of the equilibrium and near-equilibrium APG TBLs in table \ref{tab:cases}. 

The dependence of $P_0^+$ and $y_P^*$ on $\beta$ and $Re_\tau$ deserves further investigations. As indicated by the dashed line in figure \ref{fig:tau_result_E}(a), $P_0^+$ is related to the overshoot of $\tau^+$ in the case of APG. The overshoot is quantified with $\tau_{\rm{max}}^+-1$. A reasonable thinking would be $\tau_{\rm{max}}^+-1\propto P_0^+$. Therefore, according to (\ref{eq:stress_max}), $g(A)$ and thus $A$ should be independent of $\beta$ and $Re_\tau$. Consequently, $y_P^*\propto {[P_0^+/(1+P_0^+)]}^{2/3}$. 

\begin{figure}
  \begin{subfigure}[b]{0.5\linewidth}
    \centering
    \includegraphics[width=\linewidth]{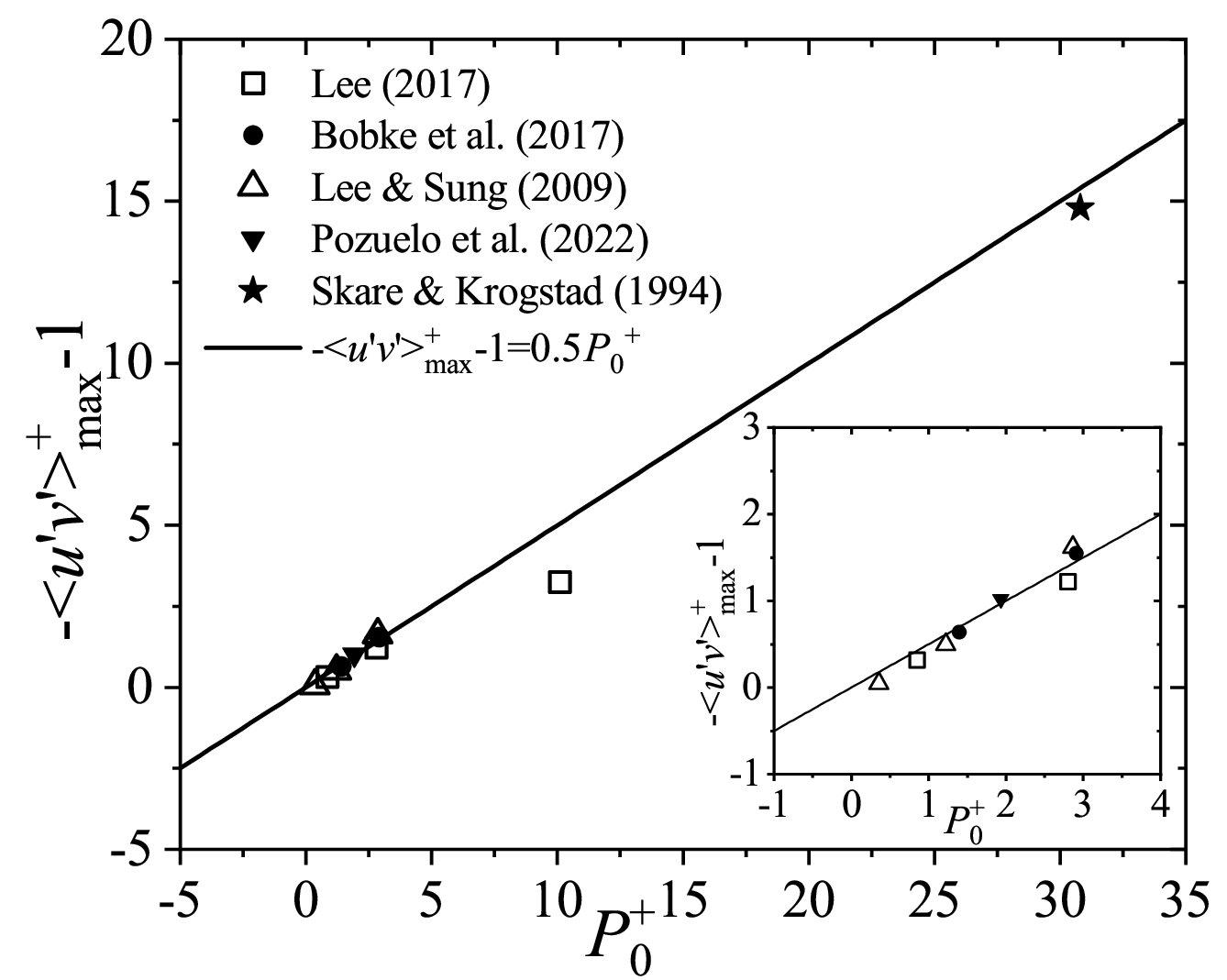}
    \caption{}
  \end{subfigure}
  \begin{subfigure}[b]{0.5\linewidth}
    \centering
    \includegraphics[width=\linewidth]{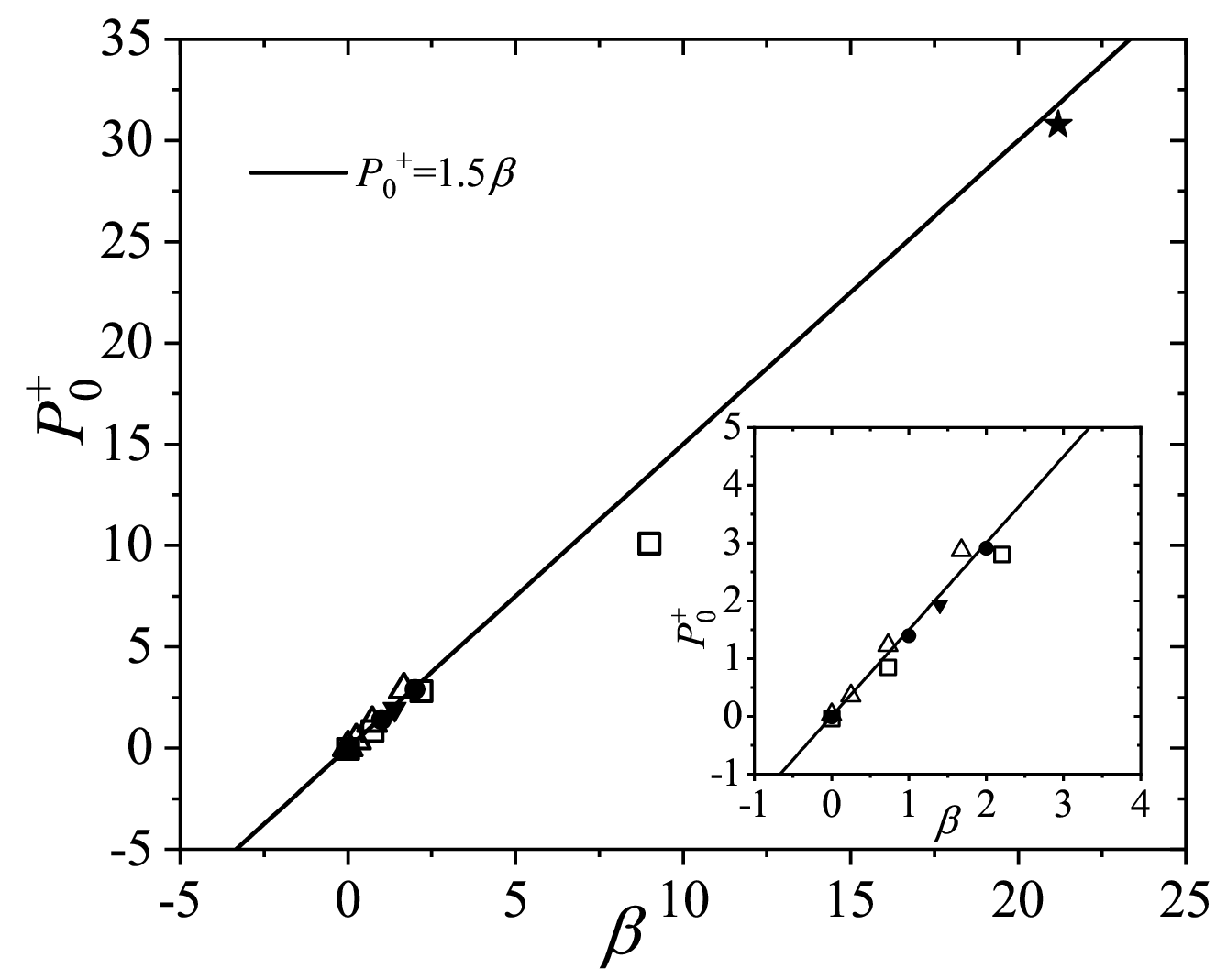}
    \caption{}
  \end{subfigure}  
%\begin{center}
  \begin{subfigure}[b]{0.5\linewidth}
    \centering
    \includegraphics[width=\linewidth]{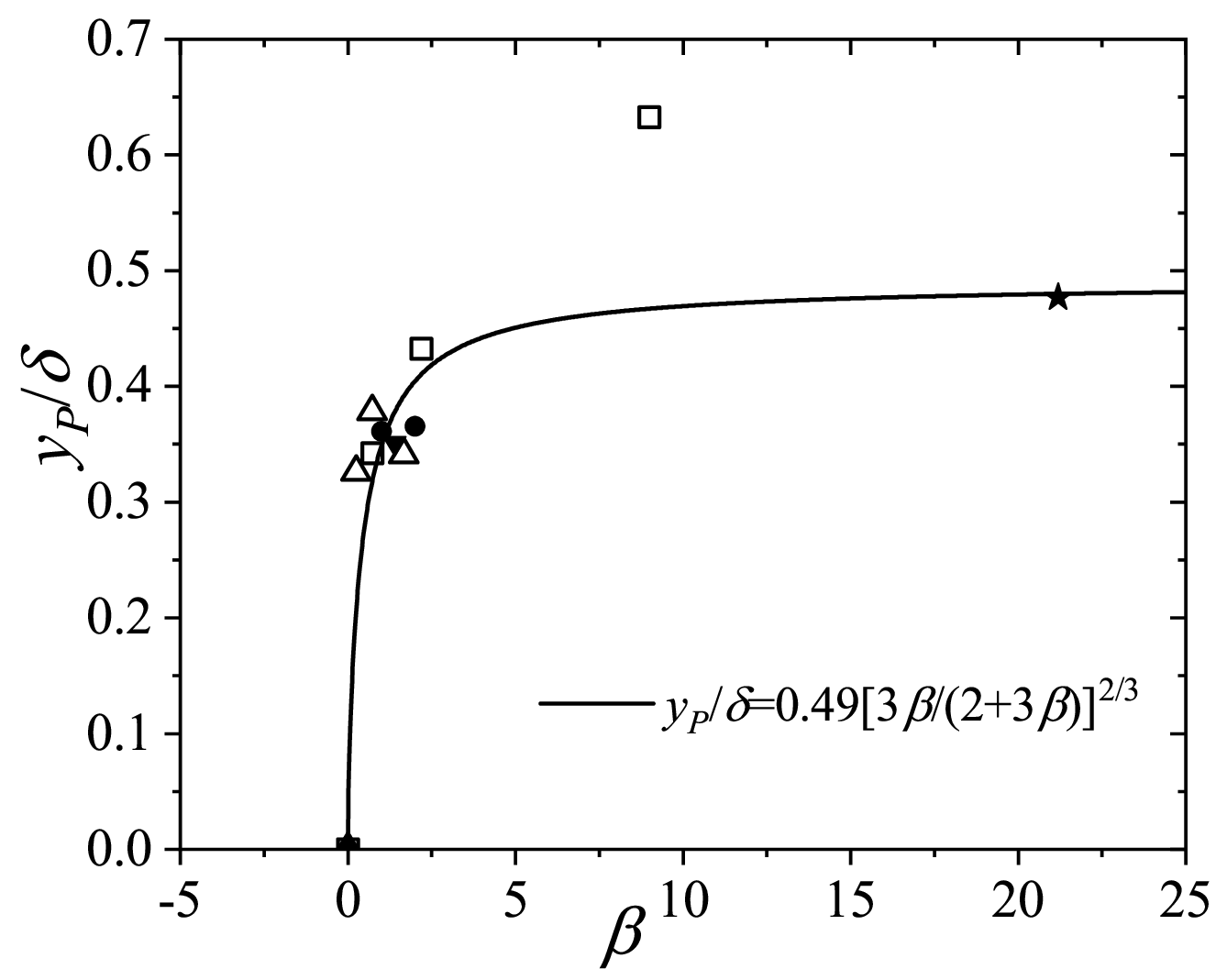}
    \caption{}
  \end{subfigure} 
%\end{center}
  \begin{subfigure}[b]{0.5\linewidth}
    \centering
    \includegraphics[width=\linewidth]{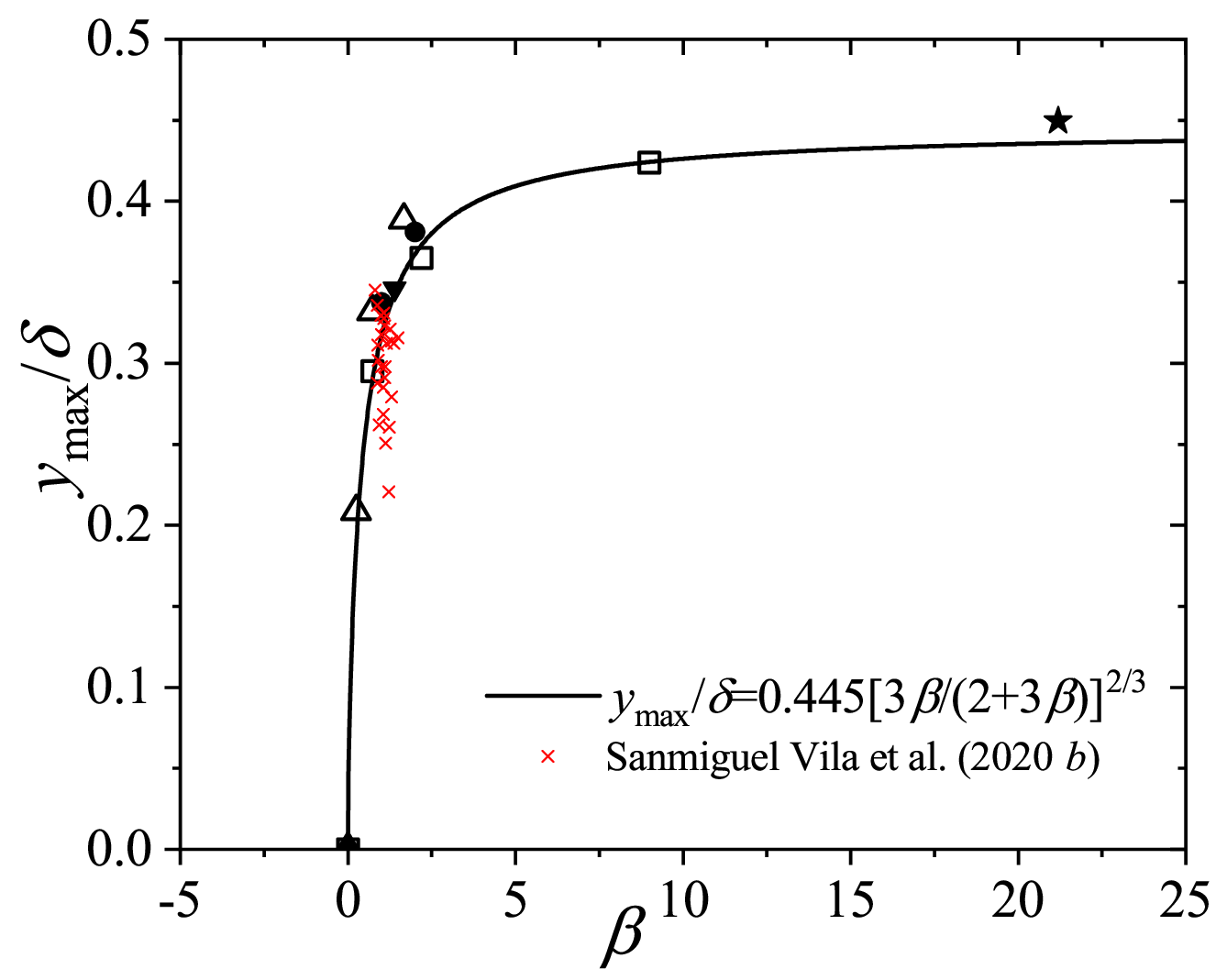}
    \caption{}
  \end{subfigure} 
   \caption{(a) $-{\left<u'v'\right>}_{\rm{max}}^+-1$ versus $P_0^+$, (b) $P_0^+$ versus $\beta$, (c) $y_P^*$ versus $\beta$, and (d) $y_{\rm{max}}^*$ versus $\beta$ for the equilibrium PG TBLs in table \ref{tab:cases}, comparing with the current models. The insets in (a) and (b) display the cases with small $\beta$.}
  \label{fig:uvmax_P0_beta}
\end{figure}

Because the peak total stress is located in the bulk of the boundary layer, $\tau_{\rm{max}}^+=-{\left<u'v'\right>}_{\rm{max}}^+$. Figure \ref{fig:uvmax_P0_beta}(a) shows the correlation between $-{\left<u'v'\right>}_{\rm{max}}^+-1$ and $P_0^+$, both being empirically measured with the studied databases. Indeed, for the studied flows, there roughly is a proportional relationship: $-{\left<u'v'\right>}_{\rm{max}}^+-1=0.5P_0^+$. Scatterings occur only at a limited level, which is inevitable owing to the effects of finite $Re_\tau$, measurement uncertainty and difficulty in keeping equilibrium in experiment, and sensitivity to initial condition and flow history in numerical simulation \citep{devenport2022equilibrium}, etc. The $\beta=9$ case of \cite{Lee2017direct} departs apparently from the trend, possibly because of a coupling of the strong APG and the rather limited $Re_\tau$ (about 330). Actually, at such $\beta$ and $Re_\tau$, the wake region significantly encroaches the inner region, such that the inner flow is not fully developed and the TBL is untypical.  

Via fitting experimental data within $0<\beta<62$, \cite{Skare1994exp} have proposed that $-{\left<u'v'\right>}_{\rm{max}}^+=1+0.75\beta$ in equilibrium APG TBLs, independently of $Re_\tau$. As a consequence, $P_0^+$ should be proportional to $\beta$. The variation of $P_0^+$ with $\beta$ is displayed in figure \ref{fig:uvmax_P0_beta}(b) for the studied databases. For the equilibrium APG TBLs, $P_0^+$ is indeed proportional to $\beta$ over the investigated ranges of $\beta$ and $Re_\theta$, and the scatterings between different cases are only minor. Therefore, for equilibrium PG TBLs, we find that $P_0^+=1.5\beta$, which agrees with the proposal of \cite{Skare1994exp}. 

With these empirical proportional relationships, we can derive from (\ref{eq:stress_max}) that $y_P^*\approx0.49[3\beta/(2+3\beta)]^{2/3}$, which is validated in figure \ref{fig:uvmax_P0_beta}(c). The prediction agrees with the data quite well except the $\beta=9$ case of \cite{Lee2017direct}. Because $y_P^*\approx(P_0^+/\beta)(\delta_1/\delta)$, $\delta_1/\delta$ should approximately also be proportional to $[3\beta/(2+3\beta)]^{2/3}$. Consequently, $\delta_1/\delta$ becomes zero in the case of ZPG. This may not be strange, but reveals that the canonical ZPG TBL is strictly in the equilibrium state only when the Reynolds numbers are infinitely large \citep{devenport2022equilibrium}. Probably because of this reason, $y_P^*$ limitedly deviates from its theoretical value of $(P_0^+/\beta)(\delta_1/\delta)$ at finite $Re_\tau$, and should be treated as a free parameter in (\ref{eq:tau_PG}).

Furthermore, according to (\ref{eq:y_max_tau}), the wall-normal location with peak total stress is predicted by
\begin{equation}
  y_{\rm{max}}^*\approx0.455\left(\frac{3\beta}{2+3\beta}\right)^{2/3}.
  \label{eq:ymax}
\end{equation}
This crucial prediction is validated in figure \ref{fig:uvmax_P0_beta}(d) with the databases in table \ref{tab:cases} as well as a recent experimental dataset by \cite{Sanmiguel2020PRF}. It is shown that $y_{\rm{max}}^*$ increases rapidly with increasing $\beta$ when $\beta$ is less than about 2.5, but quickly approaches a constant (about 0.445) when $\beta$ is larger. Aware that $y_{\rm{max}}^*$ depends on $\beta$, \cite{wei_knopp_2023} recently suggested that $y_{\rm{max}}\approx0.4\delta$ could be seen as a low-order approximation. Our prediction presents a higher-order correction for their estimation by taking into account the apparent PG effect on $y_{\rm{max}}$. It is interesting that the $\beta=9$ case of \cite{Lee2017direct} accurately agrees with (\ref{eq:ymax}), reavealing that $y_{\rm{max}}^*$ is much less sensitive to the finite-Reynolds-number effect than the magnitude of the peak stress.

\begin{figure}
    \centering
    \includegraphics[width=0.5\linewidth]{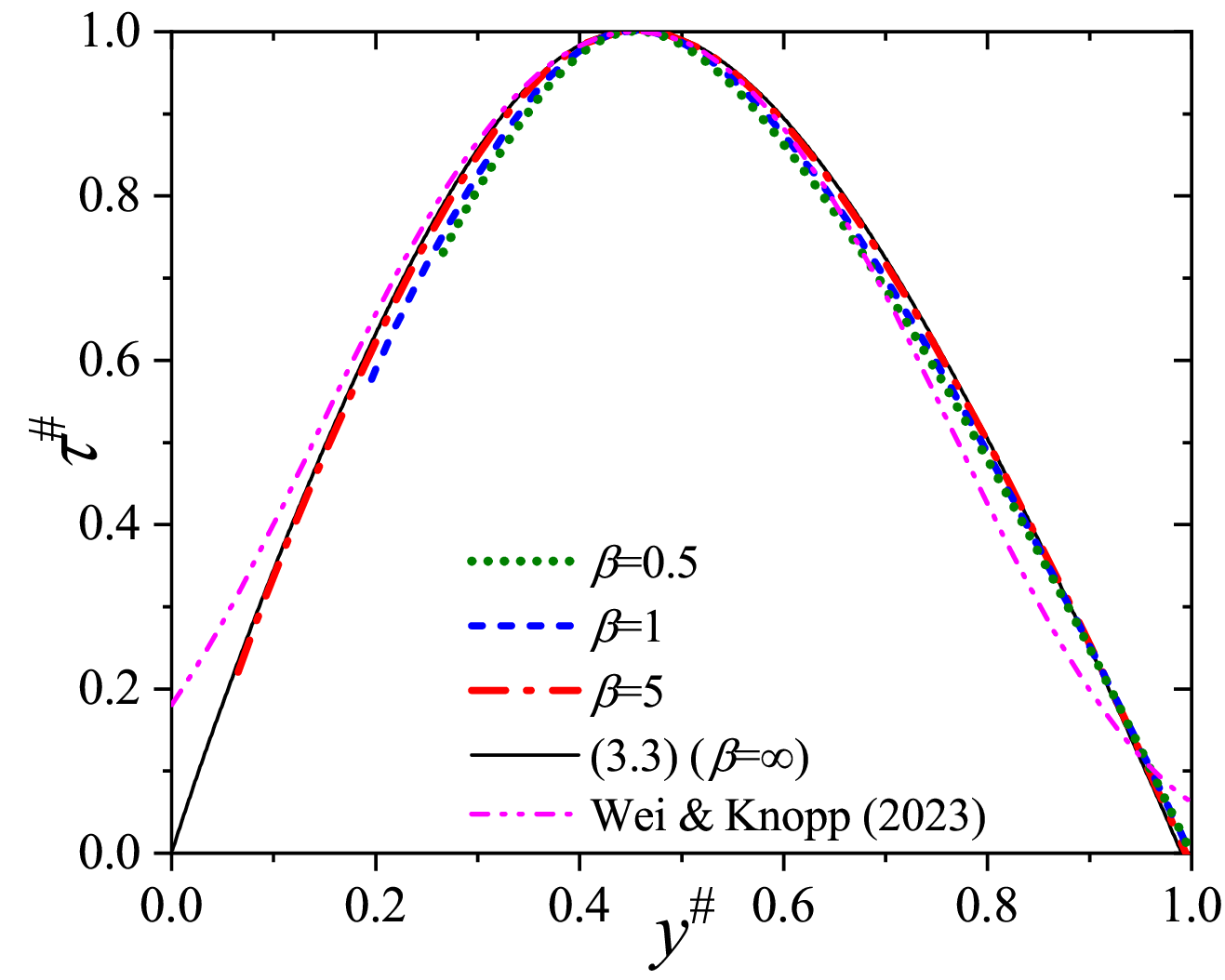}
  \caption{Validation of the transformation (\ref{eq:tau_transform}) for different $\beta$, compared with the profile of \cite{wei_knopp_2023}.}
  \label{fig:transformation_tau}
\end{figure}

With $P_0^+=1.5\beta$ and $y_P^*=0.49[3\beta/(2+3\beta)]^{2/3}$ in (\ref{eq:tau_PG}), we derive a complete model for $\tau^+$. In the case of $\beta$ being sufficiently large (e.g. $\beta>10$), $\tau^+/\tau_{\rm max}^+$ approaches an ultimate state: $\tau^+/\tau_{\rm max}^+=2\left\{{\left[ {1 + {{\left( {0.49/y^*} \right)}^4}} \right]}^{-1/4}-{y^*}^{3/2}\right\}$. It is interesting that the $\tau^+$ profiles at smaller $\beta$ can approximately be transformed into the ultimate state via the following transformation:
\begin{equation}
\left. \begin{array}{ll}
\displaystyle \tau^\#=\frac{\tau^+}{\tau_{\rm max}^+},\quad \quad \\[8pt]
\displaystyle  y^\#=1-0.545\frac{1-y^*}{1-y_{\rm max}^*},\quad \quad 
 \end{array}\right\}
  \label{eq:tau_transform}
\end{equation}
where $\tau^\#$ possesses the ultimate-state profile:
\begin{equation}
  \tau^{\#}=2\left\{{\left[ {1 + {{\left( {\frac{0.49}{y^\#}} \right)}^4}} \right]}^{-1/4}-{y^\#}^{3/2}\right\}.
  \label{eq:tau_infinite_beta}
\end{equation}
The transformation is validated for several $\beta$ in figure \ref{fig:transformation_tau} with $\tau^+$ computed by the theoretical model, i.e. (\ref{eq:tau_PG}), and with $y_{\rm max}^*$ computed by (\ref{eq:ymax}). The transformation works quite well for small and moderate $\beta$. It should be pointed out that, for a small $\beta$, the transformed profile is located in the spatial domain of $1-0.545/(1-y_{\rm max}^*)\leq y^\#\leq1$ with $1-0.545/(1-y_{\rm max}^*)>0$.

Note that our transformation is equivalent to the outer-layer scaling of the Reynolds shear stress proposed by \cite{wei_knopp_2023}. In their proposal $y^*$ is transformed to $y^{**}$ with $y^{**}=(y^*-y_{\rm{max}}^*)/(1-y_{\rm{max}}^*)$, such that the peak Reynolds shear stress is located at $y^{**}=0$. In comparison, in our transformation the peak Reynolds shear stress is located at $y^\#=0.455$. With $y^{**}=(y^\#-0.455)/0.545$, the invariant Reynolds shear stress profile, i.e. $\tau^+/\tau_{\rm max}^+=exp(-{(1.3y^{**})}^2-0.385{(1.3y^{**})}^4)$, proposed by \cite{wei_knopp_2023} via data fitting, can be expressed with the $y^\#$ coordinate. As plotted in figure \ref{fig:transformation_tau}, the invariant profile of \cite{wei_knopp_2023} is in good agreement with our expression of the ultimate-state profile.

Further validation of the transformation by using experimental and numerical data is conducted in the next subsection.

\subsection{The Reynolds shear and normal stresses}\label{subsec:Reynolds_stress_profile}
\begin{figure}
  \begin{subfigure}[b]{0.5\linewidth}
    \centering
    \includegraphics[width=\linewidth]{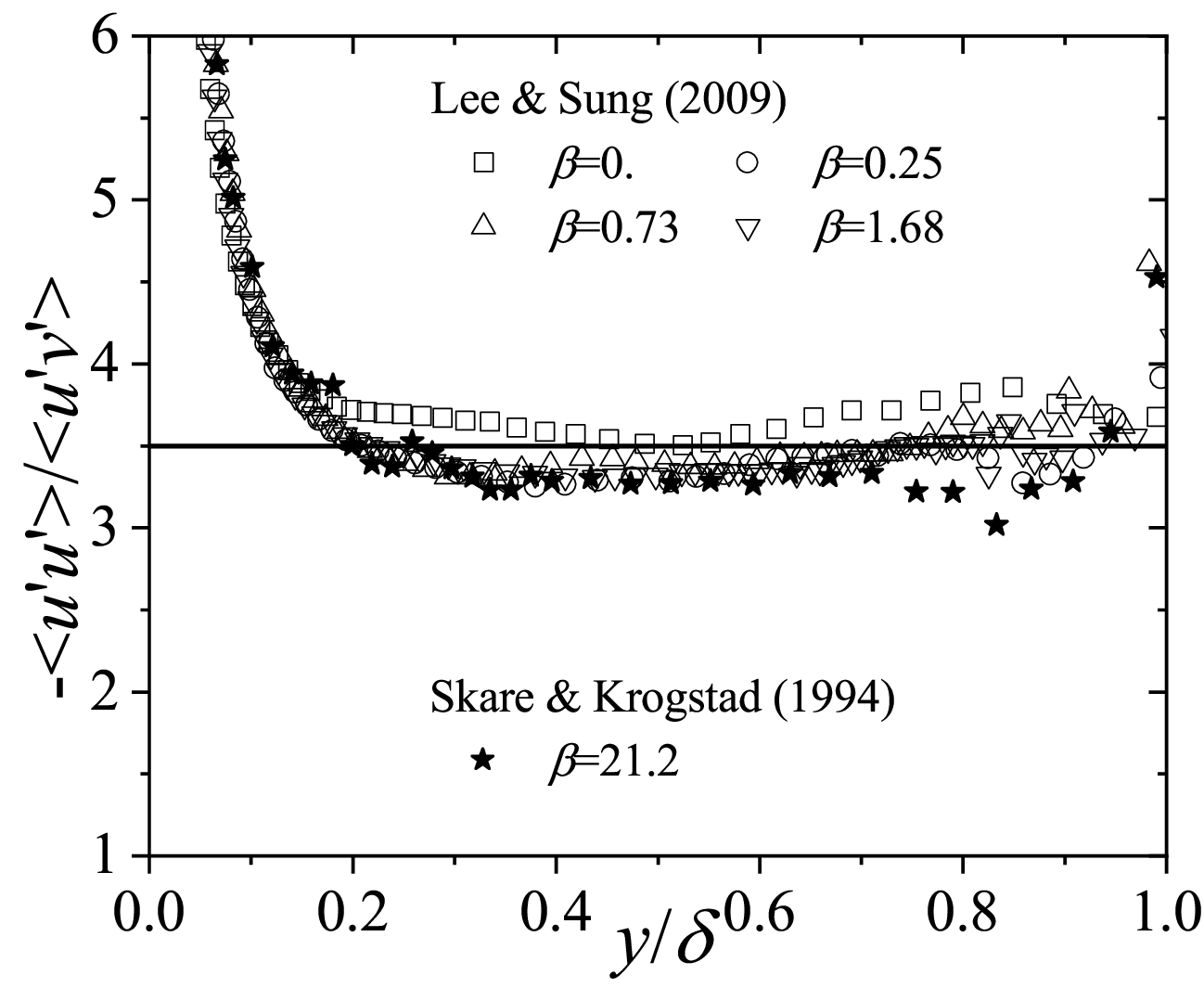}
    \caption{}
  \end{subfigure}
  \begin{subfigure}[b]{0.5\linewidth}
    \centering
    \includegraphics[width=\linewidth]{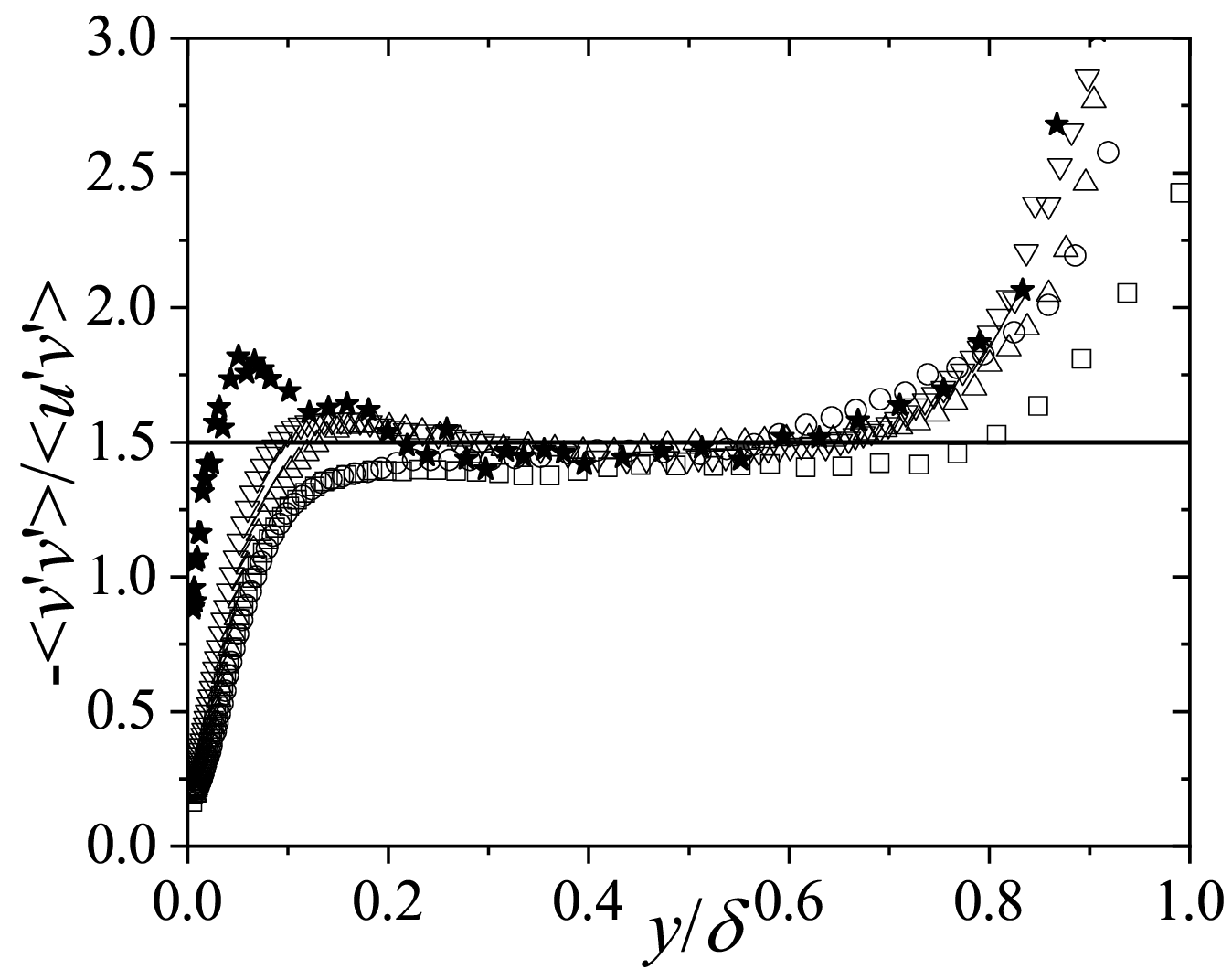}
    \caption{}
  \end{subfigure}
\begin{center}
  \begin{subfigure}[b]{0.5\linewidth}
    \centering
    \includegraphics[width=\linewidth]{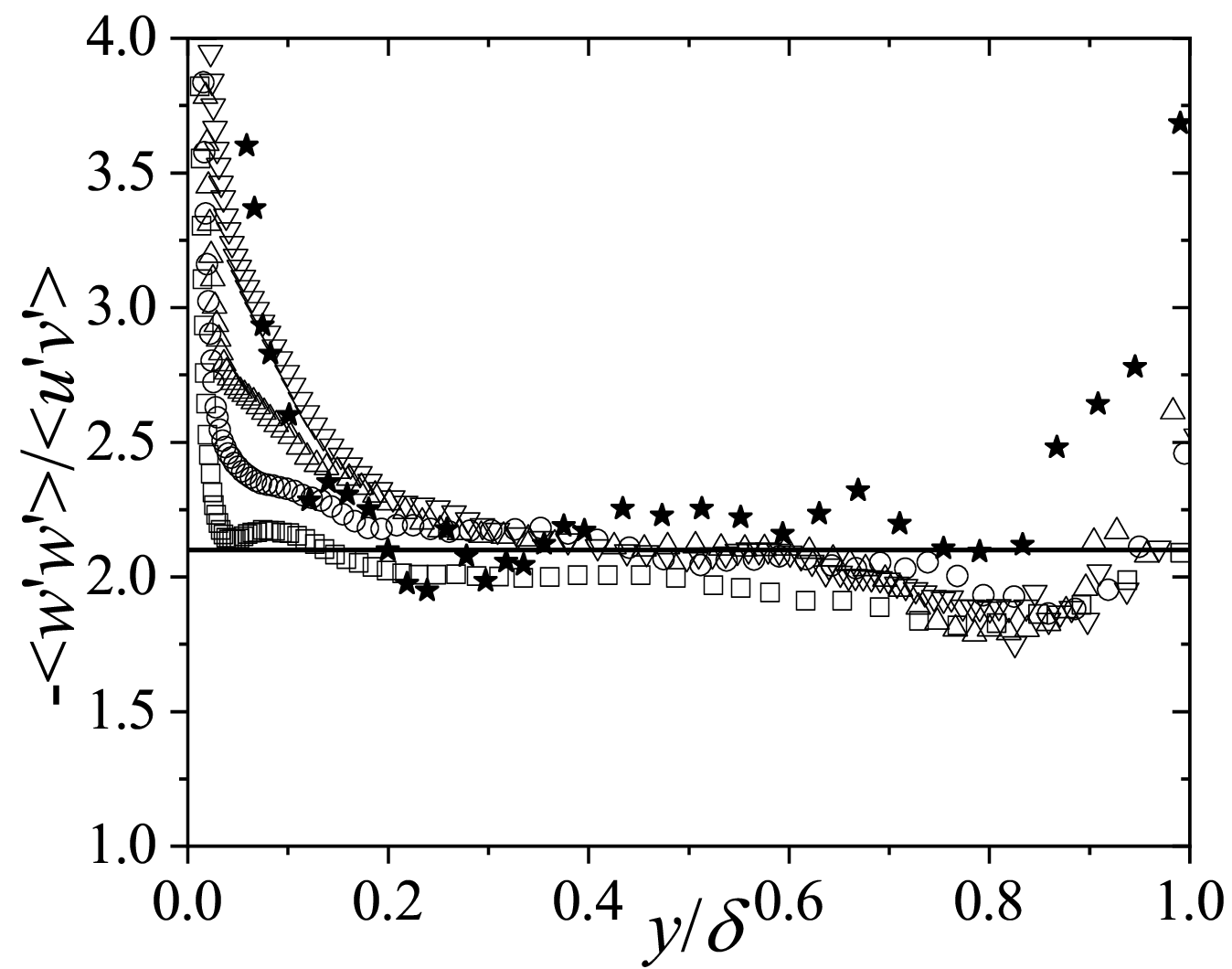}
    \caption{}
  \end{subfigure} 
\end{center}
  \caption{Profiles of the Reynolds stress ratios: (a) $-{\left<u'u'\right>}^+/{\left<u'v'\right>}^+$, (b) $-{\left<v'v'\right>}^+/{\left<u'v'\right>}^+$, and (c) $-{\left<w'w'\right>}^+/{\left<u'v'\right>}^+$. The solid line in (c) denotes $-{\left<w'w'\right>}^+/{\left<u'v'\right>}^+=2.1$.}
  \label{fig:stress_ratios}
\end{figure}

Before going to the mean velocity profiles, let us discuss the Reynolds shear and normal stresses first for the following reason. In the wake region, $-{\left<u'v'\right>}^+\approx\tau^+$, and $S^+\approx\sqrt{\tau^+}/\ell_{12}^+$. Therefore, according to (\ref{eq:kinetic_energy}) and (\ref{eq:length_out}), we can derive
\refstepcounter{equation}
$$
  {\left<u'u'\right>}^+=\frac{{\kappa_{11}}^2}{{\kappa_{12}}^2} \tau^+,\quad
  {\left<v'v'\right>}^+=\frac{{\kappa_{11}}^2}{{\kappa_{12}}^2} \tau^+,\quad
  {\left<w'w'\right>}^+=\frac{{\kappa_{11}}^2}{{\kappa_{12}}^2} \tau^+.
  \eqno{(\theequation{\mathit{a},\mathit{b},\mathit{c}})}\label{eq:kinetic_energy_wake}
$$
(\ref{eq:kinetic_energy_wake}) reveals an outer-layer similarity among the Reynolds shear and normal stresses. The validation of (\ref{eq:kinetic_energy_wake}) is shown in figure \ref{fig:stress_ratios} via the Reynolds stress ratios. To be clear, only the DNS data of \cite{Lee2009} and the experimental data of \cite{Skare1994exp} are displayed, which cover wide ranges of $\beta$ and $Re_\tau$. Similar with the observations of \cite{Lozano-Duran2019JFM} and \cite{Rkein2023dns}, the Reynolds stress ratios are nearly constant within $0.2<y^*<0.8$, almost independently of $Re_\tau$ and $\beta$. The constants, which are ${(\kappa_{11}/\kappa_{12})}^2\approx3.5$, ${(\kappa_{22}/\kappa_{12})}^2\approx1.5$, and ${(\kappa_{33}/\kappa_{12})}^2\approx2.1$, for the streamwise, wall-normal, and spanwise Reynolds stress ratios, respectively, reveal an invariant energy partition in the outer layer among different Reynolds normal stress components. Accordingly, Townsend's structure parameter $a_1\equiv-{\left<u'v'\right>}^+/({\left<u'u'\right>}^++{\left<v'v'\right>}^++{\left<w'w'\right>}^+)={\kappa_{12}}^2/({\kappa_{11}}^2+{\kappa_{22}}^2+{\kappa_{33}}^2)$ is about 0.14 in the outer region for the current databases, fairly close to that of the ZPG TBL (which is widely accepted to be 0.15). Near the boundary layer edge, however, there seems to be a tendency of equal energy partition, which is not described by the current model with constant $\kappa_{ij}$. Because the Reynolds stresses are rather small near the boundary layer edge, this defect is not significant. 

\begin{figure}
  \begin{subfigure}[b]{0.5\linewidth}
    \centering
    \includegraphics[width=\linewidth]{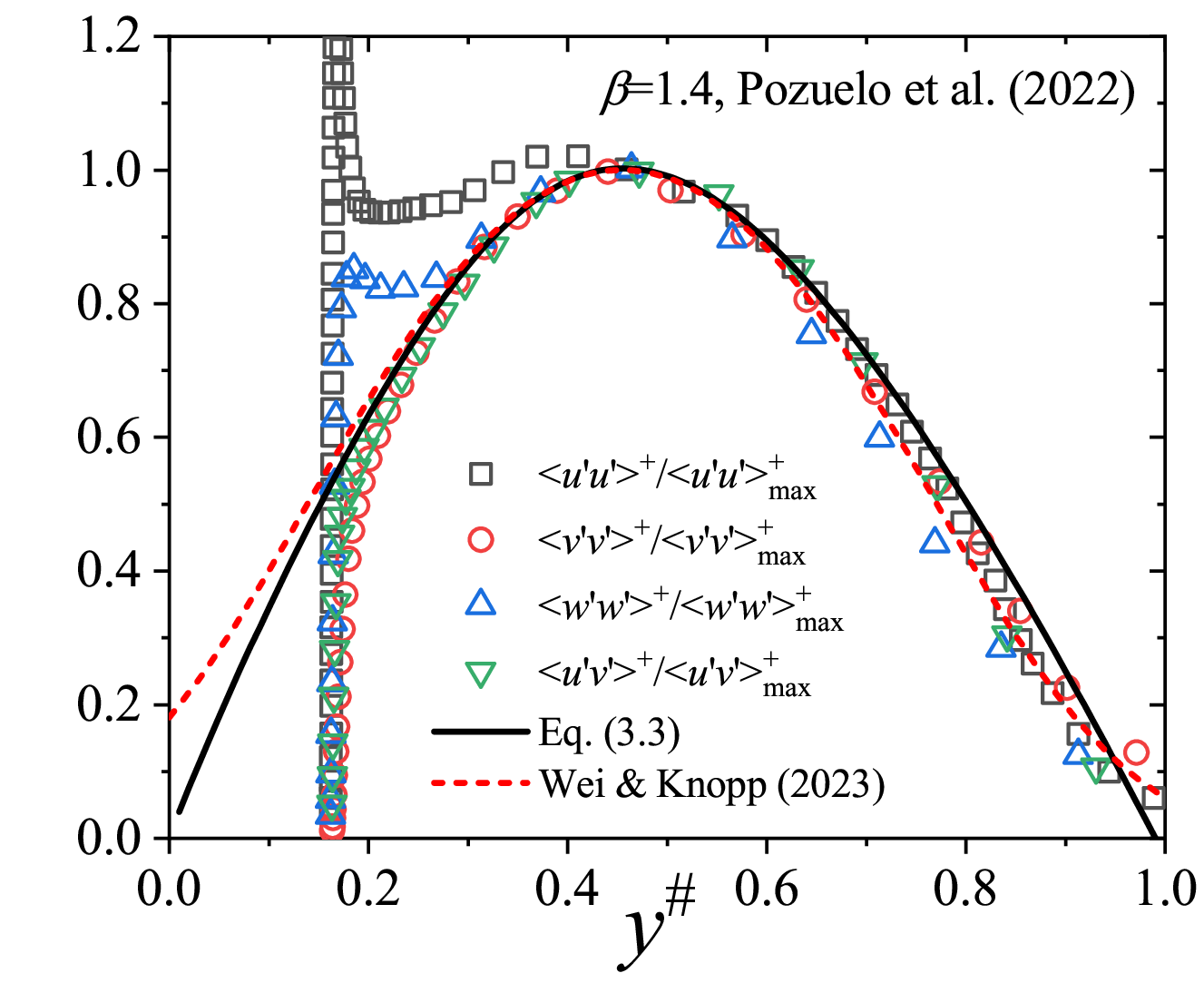}
    \caption{}
  \end{subfigure}
  \begin{subfigure}[b]{0.5\linewidth}
    \centering
    \includegraphics[width=\linewidth]{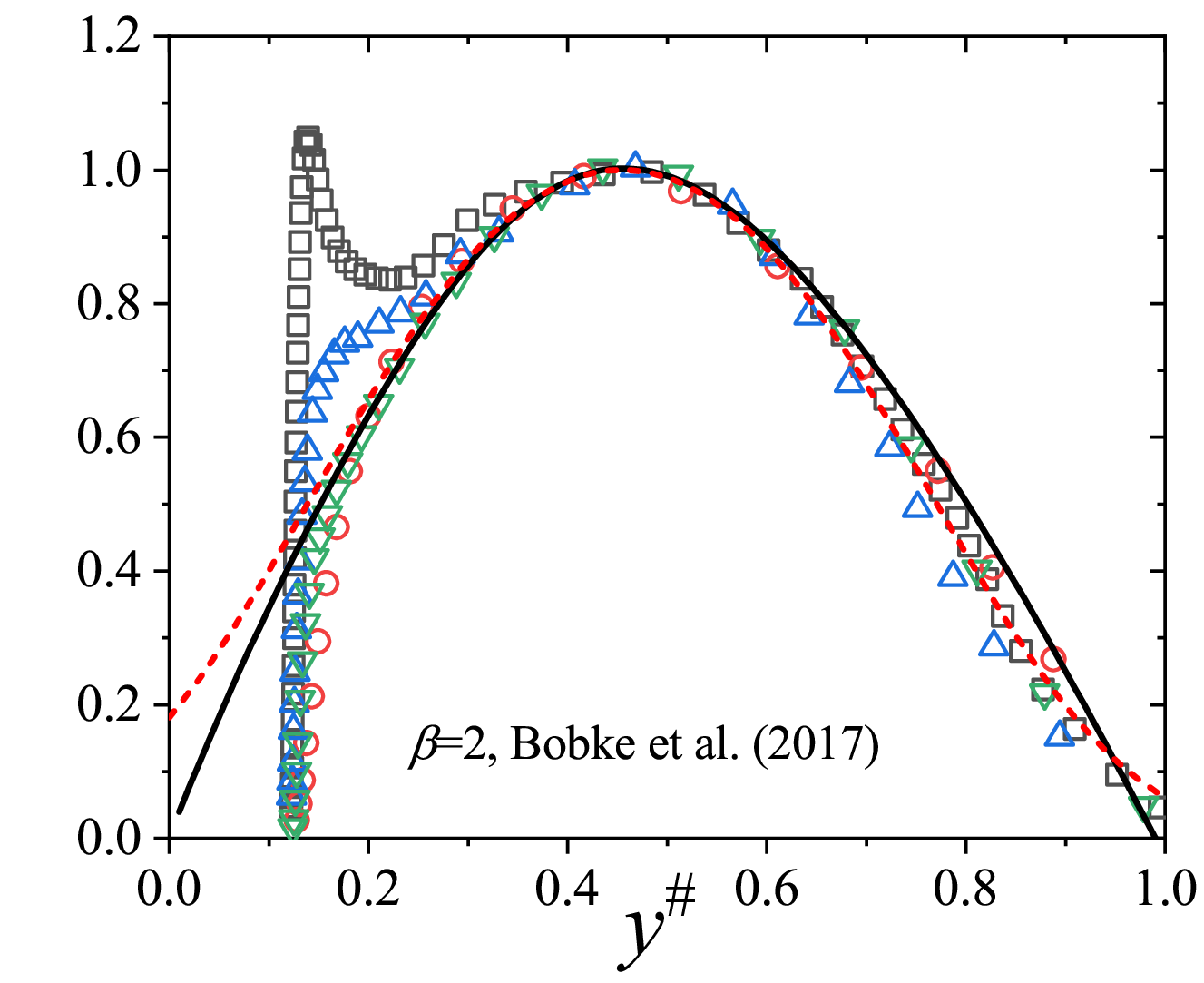}
    \caption{}
  \end{subfigure}
  \begin{subfigure}[b]{0.5\linewidth}
    \centering
    \includegraphics[width=\linewidth]{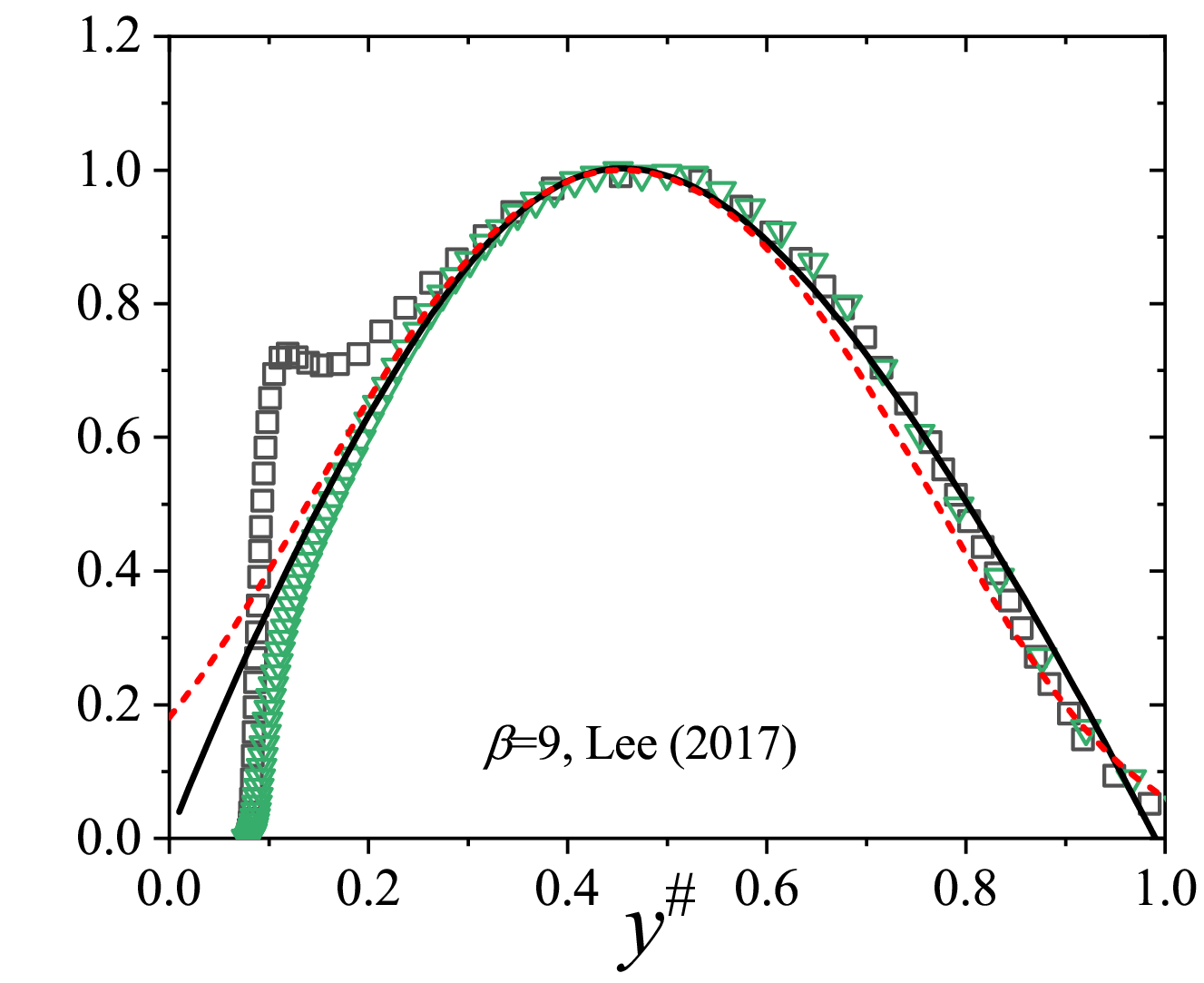}
    \caption{}
  \end{subfigure}
  \begin{subfigure}[b]{0.5\linewidth}
    \centering
    \includegraphics[width=\linewidth]{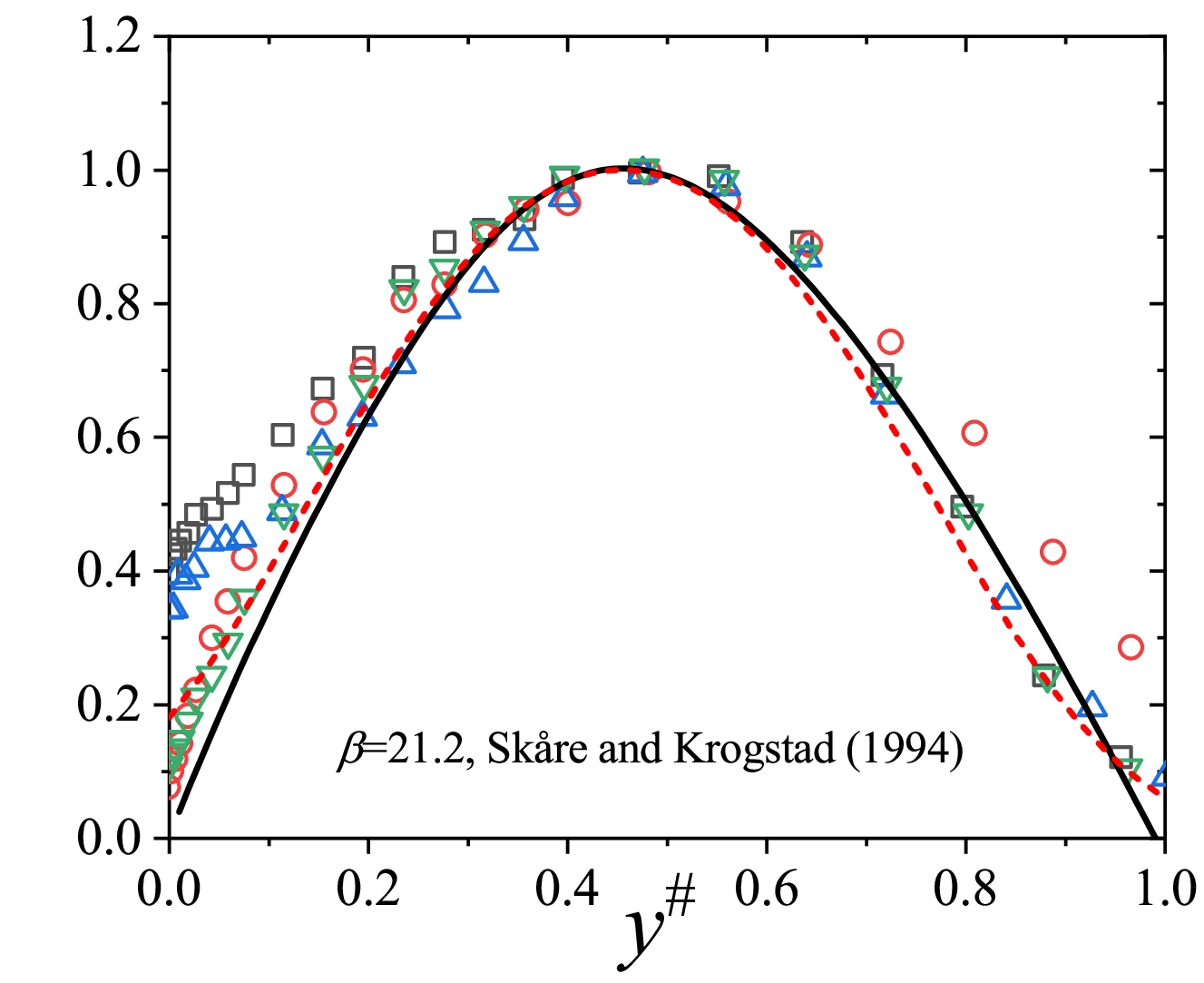}
    \caption{}
  \end{subfigure}
  \caption{Profiles of the Reynolds shear and normal stresses normalized with their corresponding peak values in the invariant coordinate of $y^\#$. Symbols denote the databases in table \ref{tab:cases}. Solid line denotes the ultimate-state profile, i.e. (\ref{eq:tau_infinite_beta}). Dashed line denotes the empirical profile of \cite{wei_knopp_2023}.}
  \label{fig:uuvvwwuv}
\end{figure}

Thus, with the constant ratios and with $\tau^+$ depicted by (\ref{eq:tau_PG}), we can predict the outer-layer profiles of the Reynolds shear and normal stresses of general equilibrium APG TBLs. Here, the validation of the prediction is performed by employing the invariant transformation (\ref{eq:tau_infinite_beta}) for the databases in table \ref{tab:cases}. To compute the invariant transformation with the data, $y_{\rm max}^*$ (same for all Reynolds stresses) and the peak Reynolds shear and normal stresses are all acquired from the databases. As shown in figure \ref{fig:uuvvwwuv}, in the wake region the data agree with the ultimate-state profile and the empirical profile of \cite{wei_knopp_2023} quite well.

It is intersting to compare the outer second peak of ${\left<u'u'\right>}^+$ of an APG TBL with that of a high-Reynolds-number ZPG TBL. In equilibirum APG TBLs, the outer second peak ${\left<u'u'\right>}_{\rm{max}}^+$ increases linearly with increasing $\beta$, nearly independently of $Re_\tau$ (if not too small). In addition, according to (\ref{eq:ymax}), the wall-normal location $y_{\rm max}^*$ is independent of $Re_\tau$, and, at large $\beta$, approaches $0.455$ which is constantly in the boundary-layer bulk. In contrast, for ZPG TBLs, the outer second peak of ${\left<u'u'\right>}^+$ appears only at rather high Reynolds numbers, and only slowly increases with increasing $Re_\tau$ \citep{Marusic2010,chen2018quantifying}. Besides, the wall-normal location of the outer second peak is propotional to $Re_\tau^{1/2}$, thus far from being in the boundary-layer bulk at high Reynolds nubmers. Furthermore, the numerical and experimental data, as well as the current theory, show that the Reynolds stress components of the equilibirum APG TBLs possess a similar profile in the wake region (figure \ref{fig:uuvvwwuv}). In the ZPG TBLs, however, both theory and experimental data reveal that there are logrithmic laws above the meso-layer for ${\left<u'u'\right>}^+$ and ${\left<w'w'\right>}^+$, but not for $-{\left<u'v'\right>}^+$ and ${\left<v'v'\right>}^+$ \citep{Townsend1976,Marusic2010,chen2018quantifying}. These differences reveal that there are different mechanisms for the outer second peak of ${\left<u'u'\right>}^+$ in APG and ZPG TBLs \citep{Sanmiguel2020PRF}. In the case of $\beta$ near one and Reynolds number sufficiently high, one may expect there are competing effects between the APG and Reynolds number on the Reynolds stresses of the APG TBL, which should make the dynamics more complicated and deserve further studies.  

\subsection{The mean velocity profile}\label{subsec:mean_velocity_profile}
\begin{figure}
    \centering
    \includegraphics[width=0.5\linewidth]{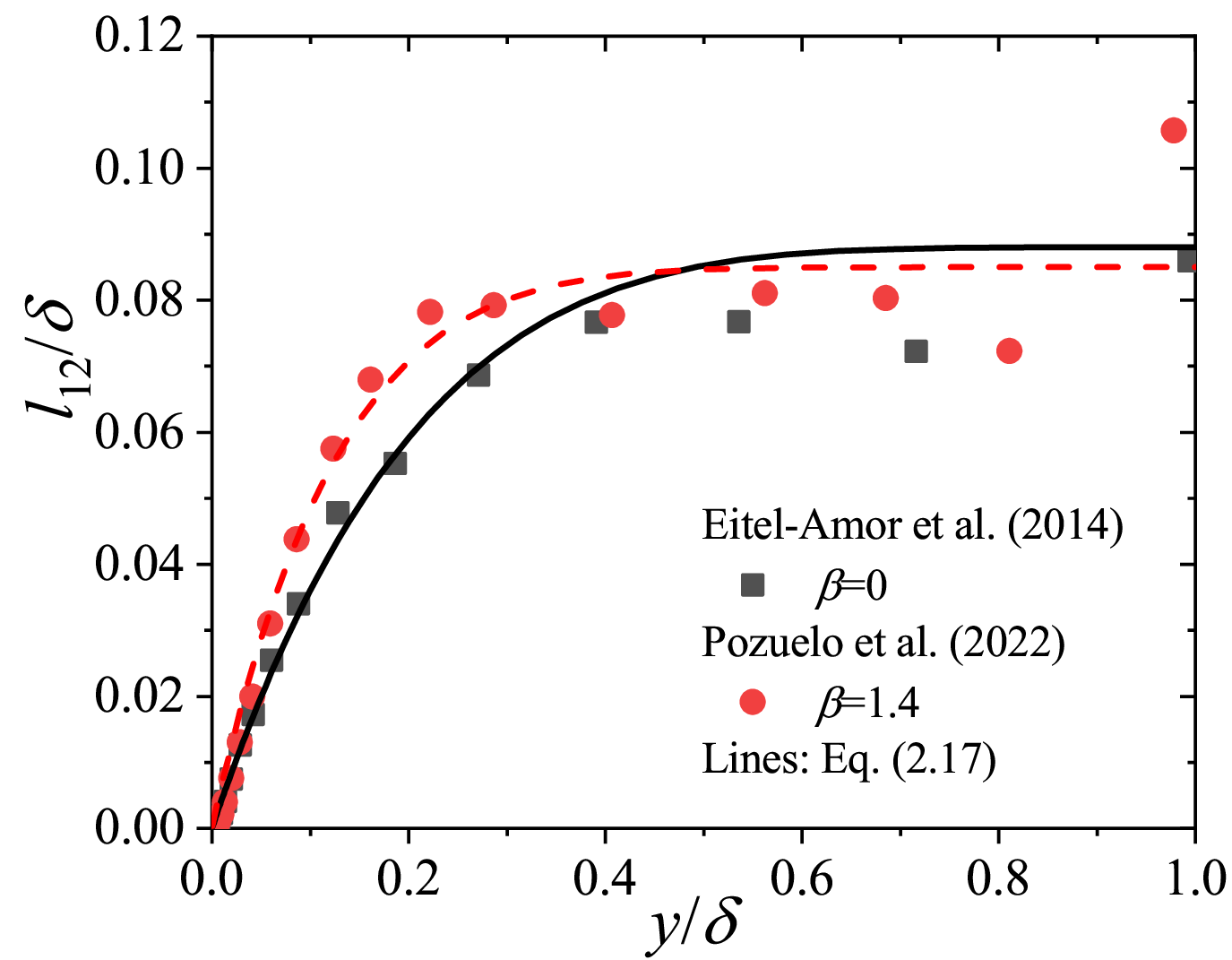}
  \caption{Comparison between the stress length model (\ref{eq:length_out}) and the LES data of \cite{Eitel-Amor2014} and \cite{Pozuelo2022les}. The red dashed line: $\kappa_{12}=0.68$, $n=8$. The black solid line: $\kappa_{12}=0.44$, $n=5$.}
  \label{fig:lm_result}
\end{figure}

The mean velocity profile in the wake region can be predicted based on the total stress and stress length models. Validation of the stress length model (\ref{eq:length_out}) is shown in figure \ref{fig:lm_result} for the ZPG TBL of \cite{Eitel-Amor2014} and the near-equilibrium APG TBL of \cite{Pozuelo2022les}, both of which have been computed with LES at matched moderate $Re_\tau$. To compute (\ref{eq:length_out}), $\kappa_{12}$ and $n$ were estimated via a least square fit of the mean velocity profile, instead of the stress length itself. In the computation, the mean velocity was predicted by (\ref{eq:udefect_model}) via (\ref{eq:tau_PG}) (with $P_0^+$ and $y_P^+$ estimated empirically as aforementioned) and (\ref{eq:length_out}), and the least square fit was conducted in the wake region of the boundary layer. We empirically set the lower boundary of the wake region at $y^+=41$, which, according to \cite{she2017quantifying}, is the buffer layer thickness (i.e. the lower boundary of the overlap layer) of ZPG TBL. Even though the buffer layer thickness may vary with PG, this setting does not markedly affect the numerical values of $\kappa_{12}$ and $n$, especially when $n$ is set as an integer. As shown in figure \ref{fig:lm_result}, (\ref{eq:length_out}) rather accurately describe the variation of $\ell_{12}$ in the wake region. Especially, the transition from the overlap layer (say, at about $y=0.15\delta$ \citep{Knopp2022}) to the constant-stress-length region is well captured by the defect power law of (\ref{eq:length_out}). Comparing with that of the ZPG TBL at matched $Re_\tau$, $\ell_{12}$ of the equilibrium APG TBL possesses a wider constant-stress-length region. The extended wake region of the APG TBL encroaches the inner flow, resulting in a steeper growth of $\ell_{12}$ beneath the wake region than that of the ZPG TBL at matched $Re_\tau$. These features are captured by (\ref{eq:length_out}) with $n$ and $\kappa_{12}$ varying with the PG as presented in the caption of figure \ref{fig:lm_result}.  

\begin{figure}
  \begin{subfigure}[b]{0.5\linewidth}
    \centering
    \includegraphics[width=\linewidth]{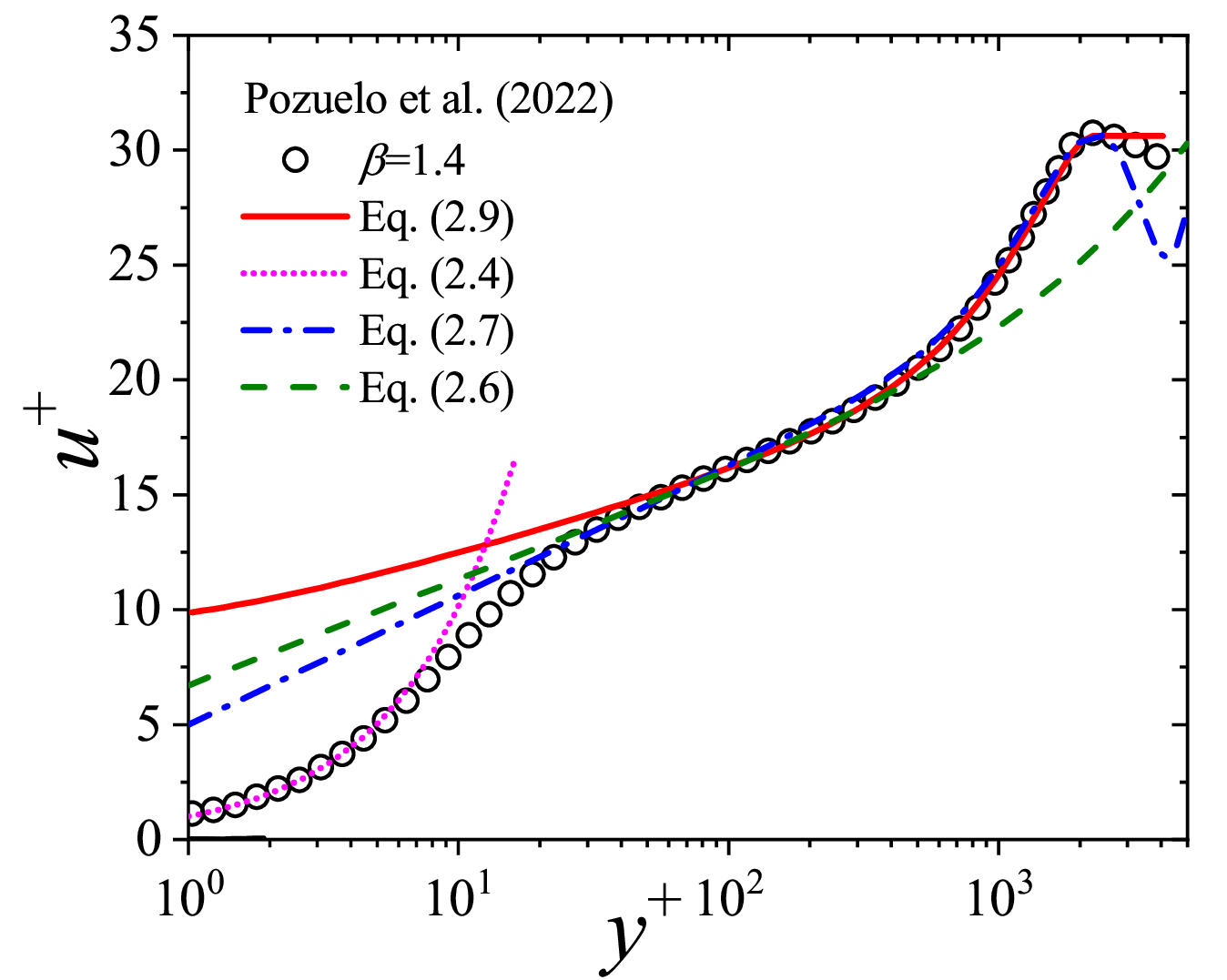}
    \caption{}
  \end{subfigure}
  \begin{subfigure}[b]{0.5\linewidth}
    \centering
    \includegraphics[width=\linewidth]{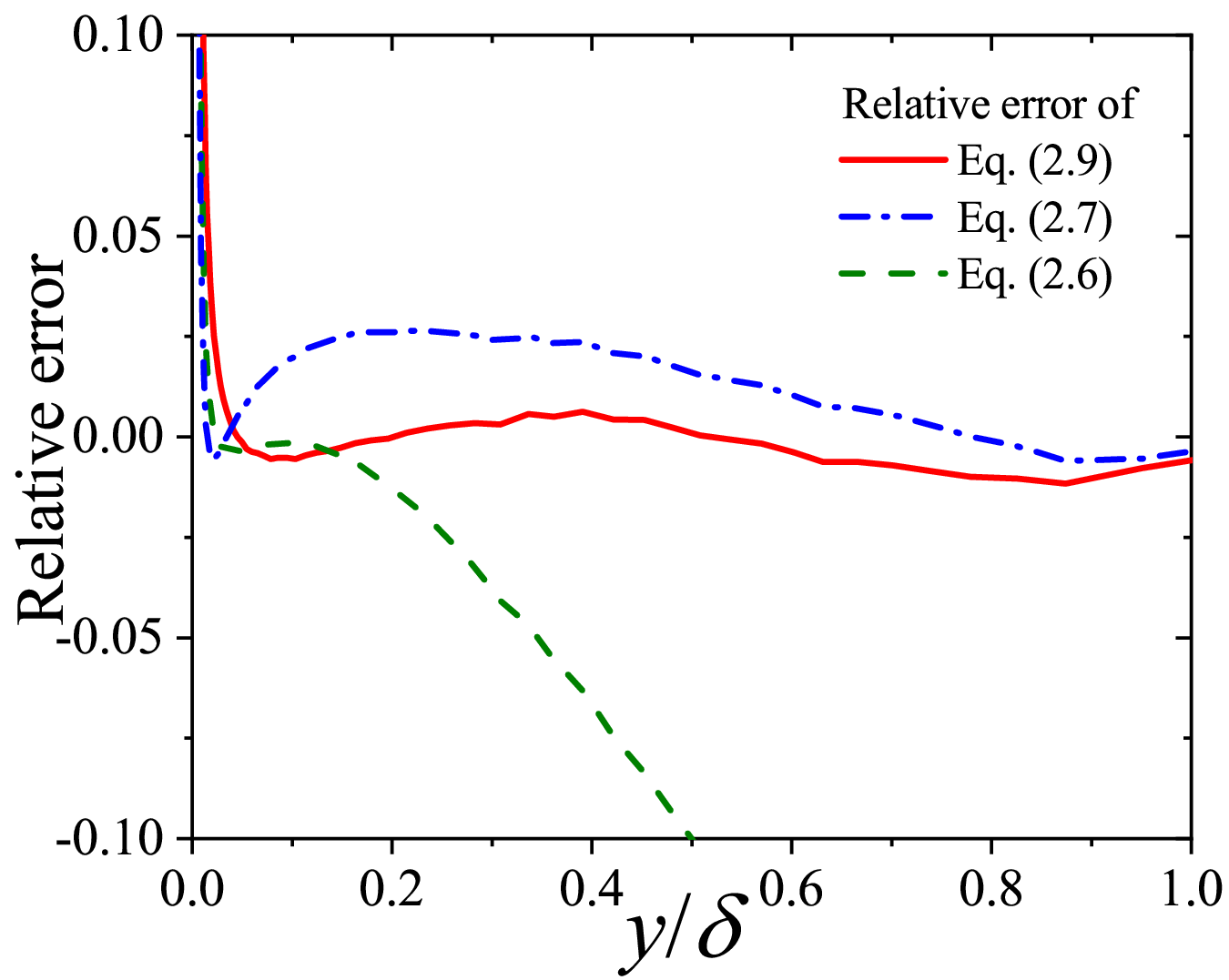}
    \caption{}
  \end{subfigure}
  \caption{(a) The streamwise mean velocity profiles predicted by different models for the $\beta=1.4$, $Re_\tau=2000$ case of \cite{Pozuelo2022les}. (b) The relative errors of the predictions. In (\ref{eq:U_nearwall}) $P_w^+=0.0031$. In (\ref{eq:U_overlap}), $\kappa_{12}=0.5$, $B=6.7$, and $\lambda=0.7P_w^+$ \citep{knopp2015investigation}. In (\ref{eq:U_wake}) $\rm{\Pi}=1.42$. To compute (\ref{eq:udefect_model}), $P_0^+=1.94$ and $y_P^+=750$ in (\ref{eq:tau_PG}), $\kappa_{12}=0.68$, $n=8$ in (\ref{eq:length_out}).}
  \label{fig:U_models}
\end{figure}

Taking the $\beta=1.4$, $Re_\tau=2000$ case of \cite{Pozuelo2022les} as an example, the $u^+$ profile predicted by (\ref{eq:udefect_model}) is compared with the predictions of the previous formulations and the LES data in figure \ref{fig:U_models}. The viscous sublayer solution (\ref{eq:U_nearwall}) and the generalized overlap law (\ref{eq:U_overlap}) describe only narrow portions of the boundary layer. The Coles' law of the wake (\ref{eq:U_wake}) rather accurately describes the $u^+$ profile within and above the overlap layer. The current outer-layer mean velocity model (\ref{eq:udefect_model}), computed with the total stress model (\ref{eq:tau_PG}) and the stress length model (\ref{eq:length_out}), presents an alternative law of the wake. As shown in figure \ref{fig:U_models}(b), whereas both describe the majority portion of the boundary layer, the current model has relative prediction errors less than about 1\% within $0.05<y^*<1$, and the Coles' law of the wake, less than about 2.5\% (which is rather good).  

\begin{figure}
  \begin{subfigure}[b]{0.5\linewidth}
    \centering
    \includegraphics[width=\linewidth]{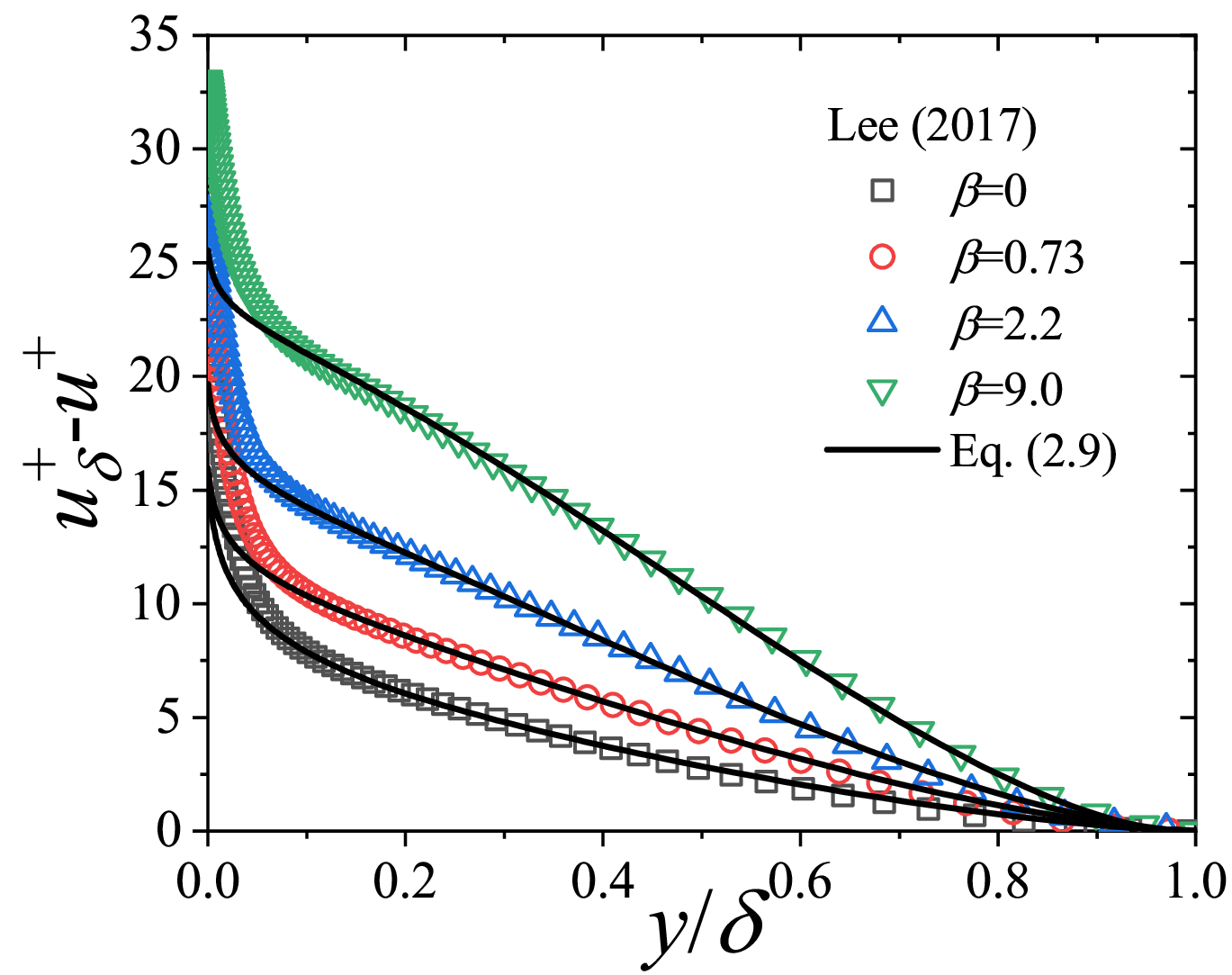}
    \caption{}
  \end{subfigure}
  \begin{subfigure}[b]{0.5\linewidth}
    \centering
    \includegraphics[width=\linewidth]{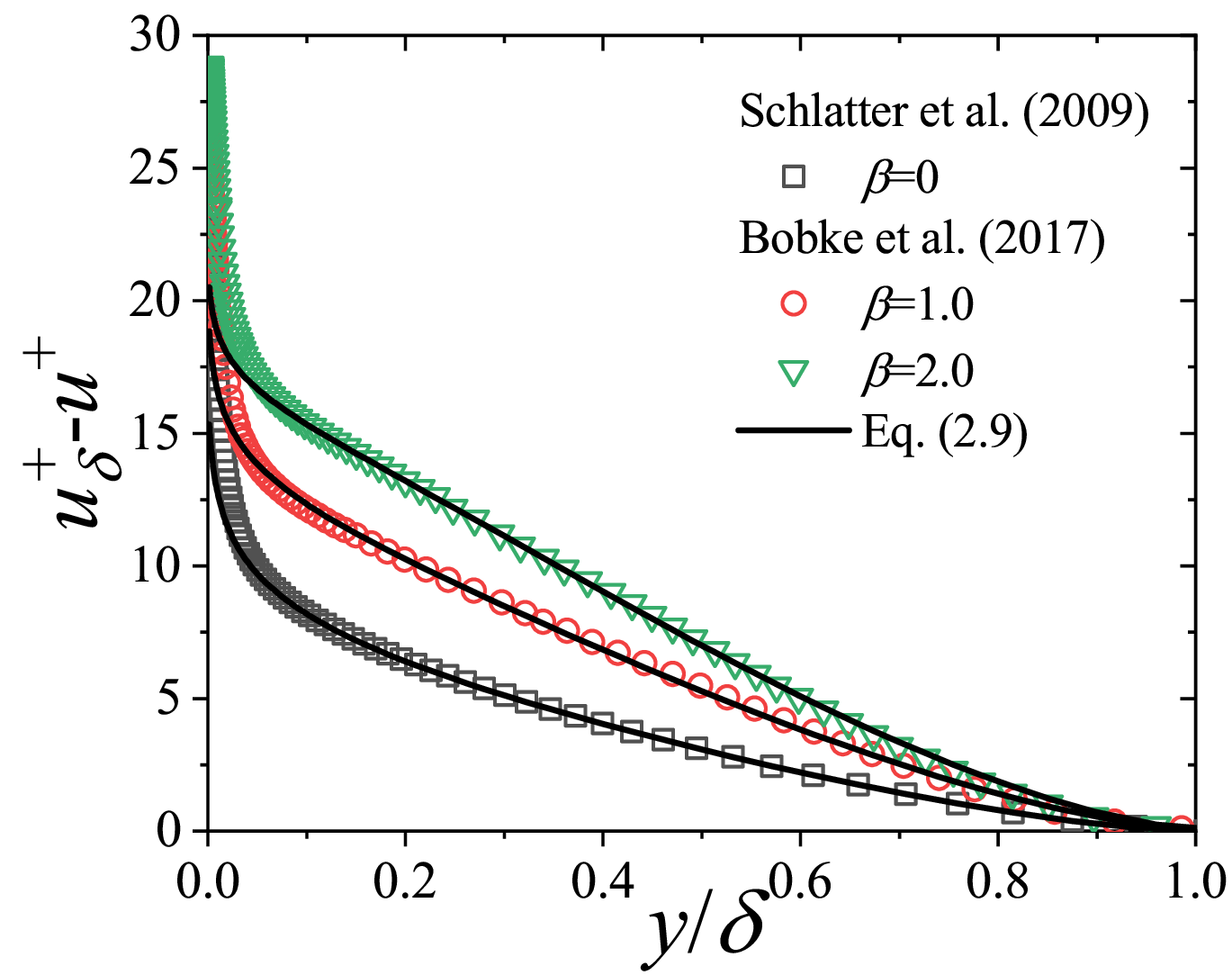}
    \caption{}
  \end{subfigure}
  \begin{subfigure}[b]{0.5\linewidth}
    \centering
    \includegraphics[width=\linewidth]{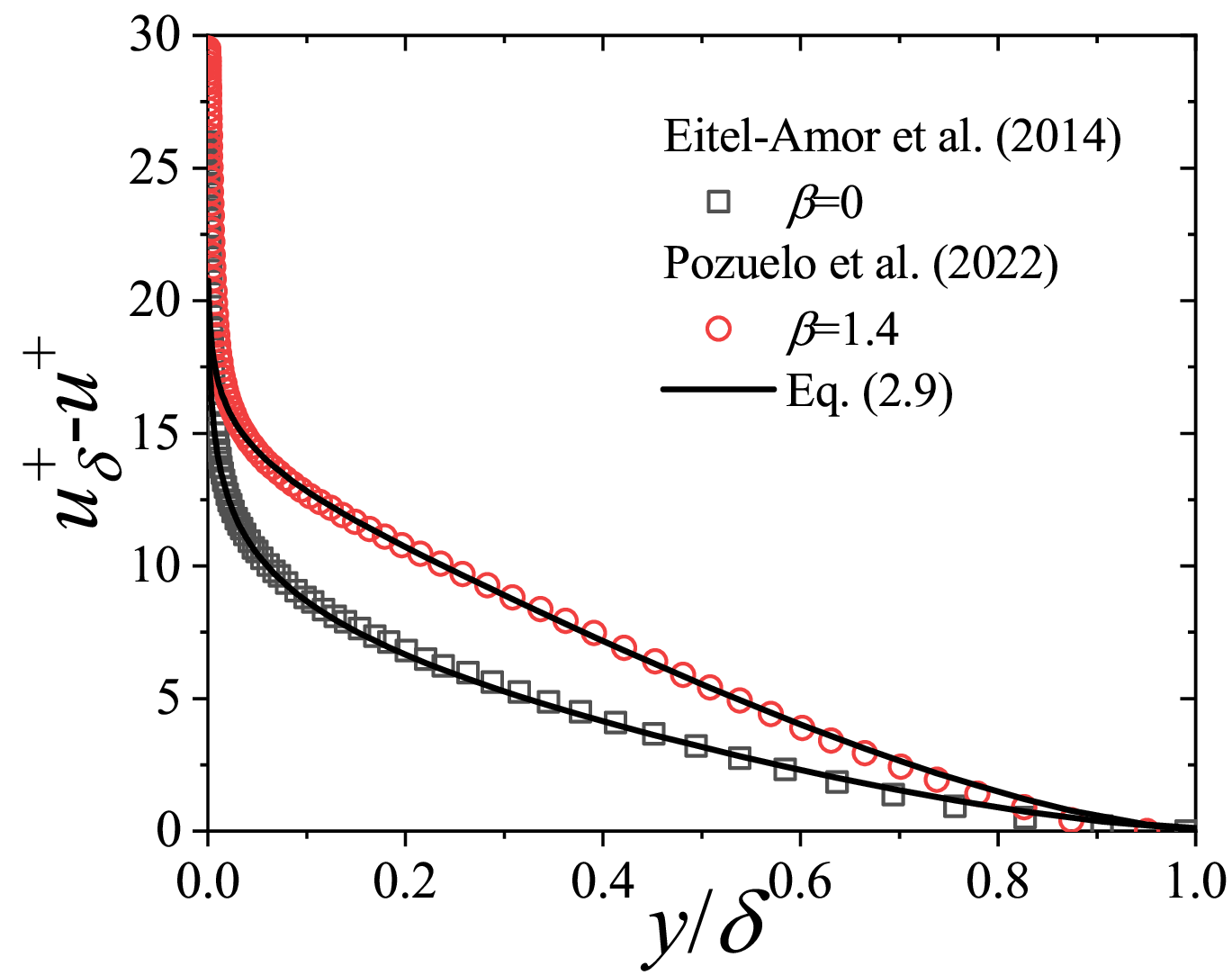}
    \caption{}
  \end{subfigure}
  \begin{subfigure}[b]{0.5\linewidth}
    \centering
    \includegraphics[width=\linewidth]{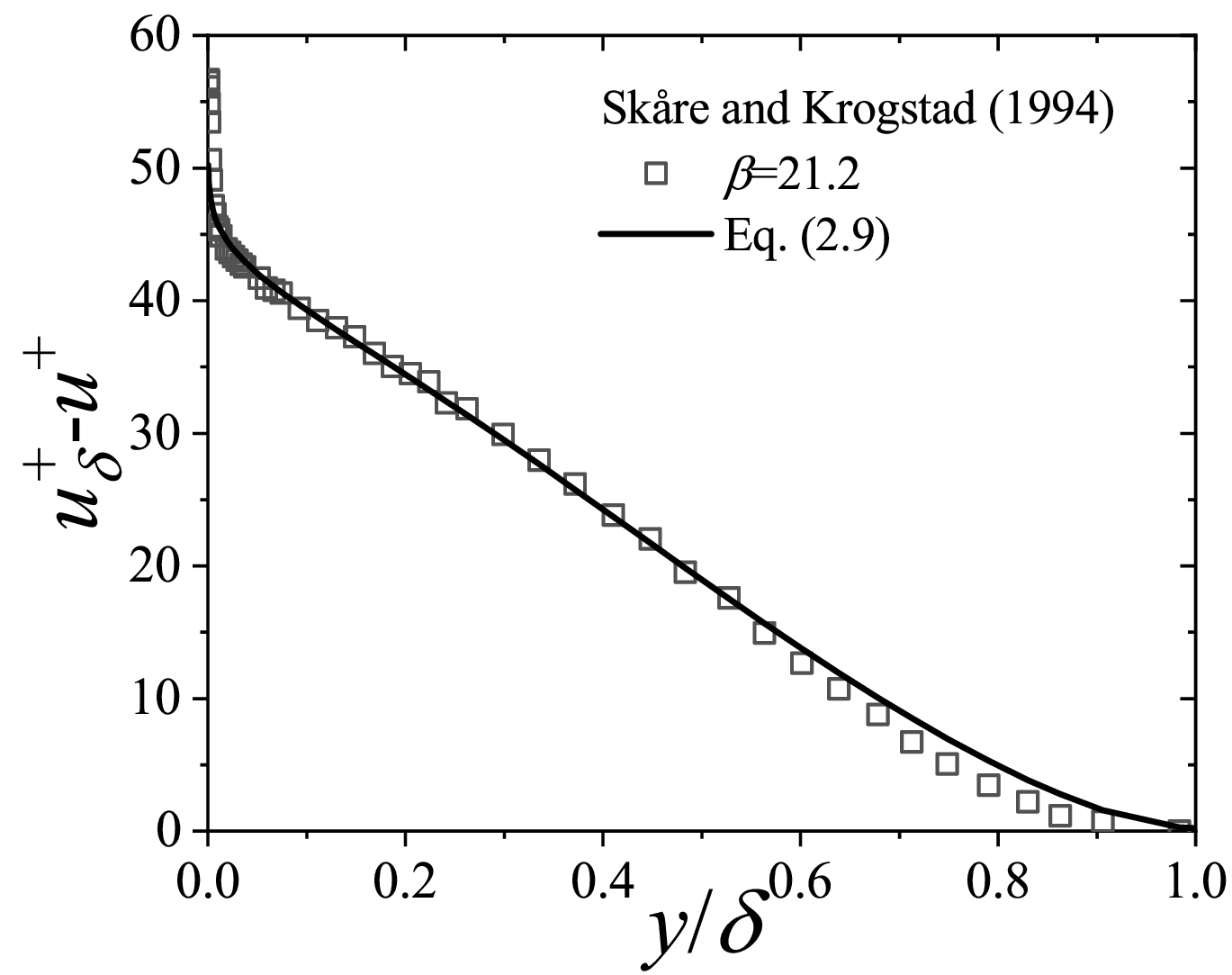}
    \caption{}
  \end{subfigure}
  \caption{The streamwise mean-velocity-deficit profiles predicted by the current model for the equilibrium PG TBLs in table \ref{tab:cases}.}
  \label{fig:U_result_E}
\end{figure}

Figure \ref{fig:U_result_E} displays the predictions of the mean-velocity-deficit profiles of the equilibrium PG TBLs in table \ref{tab:cases}. Note that, to compute (\ref{eq:udefect_model}), the parameters in (\ref{eq:tau_PG}) and (\ref{eq:length_out}) have taken the empirically-estimated values, such that the validation is posterior. As shown in figure \ref{fig:U_result_E}, the present model agrees with the data quite well over a substantial portion of the boundary layer except near the wall. In the linear coordinate, the velocity-deficit profile becomes more and more bulging in the wake region when $\beta$ is increased, which corresponds to the mean velocity profile being more and more inflected in the semi-logarithmic coordinate. This inflected mean velocity profile is one of the most prominent and intriguing features of the APG TBLs, and is accurately captured by the current model for all $\beta$ and $Re_\theta$.     

\begin{figure}
  \begin{subfigure}[b]{0.5\linewidth}
    \centering
    \includegraphics[width=\linewidth]{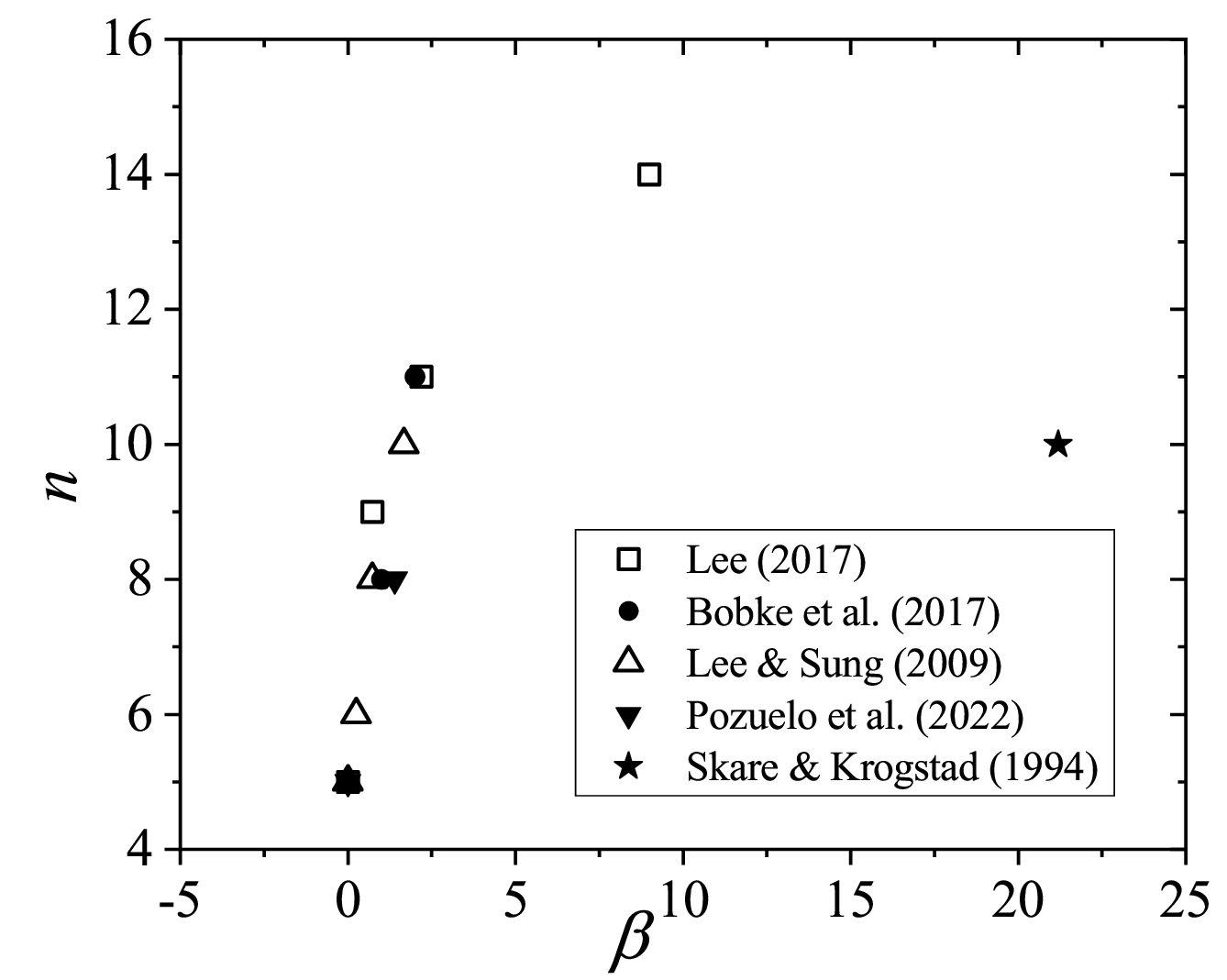}
    \caption{}
  \end{subfigure}
  \begin{subfigure}[b]{0.5\linewidth}
    \centering
    \includegraphics[width=\linewidth]{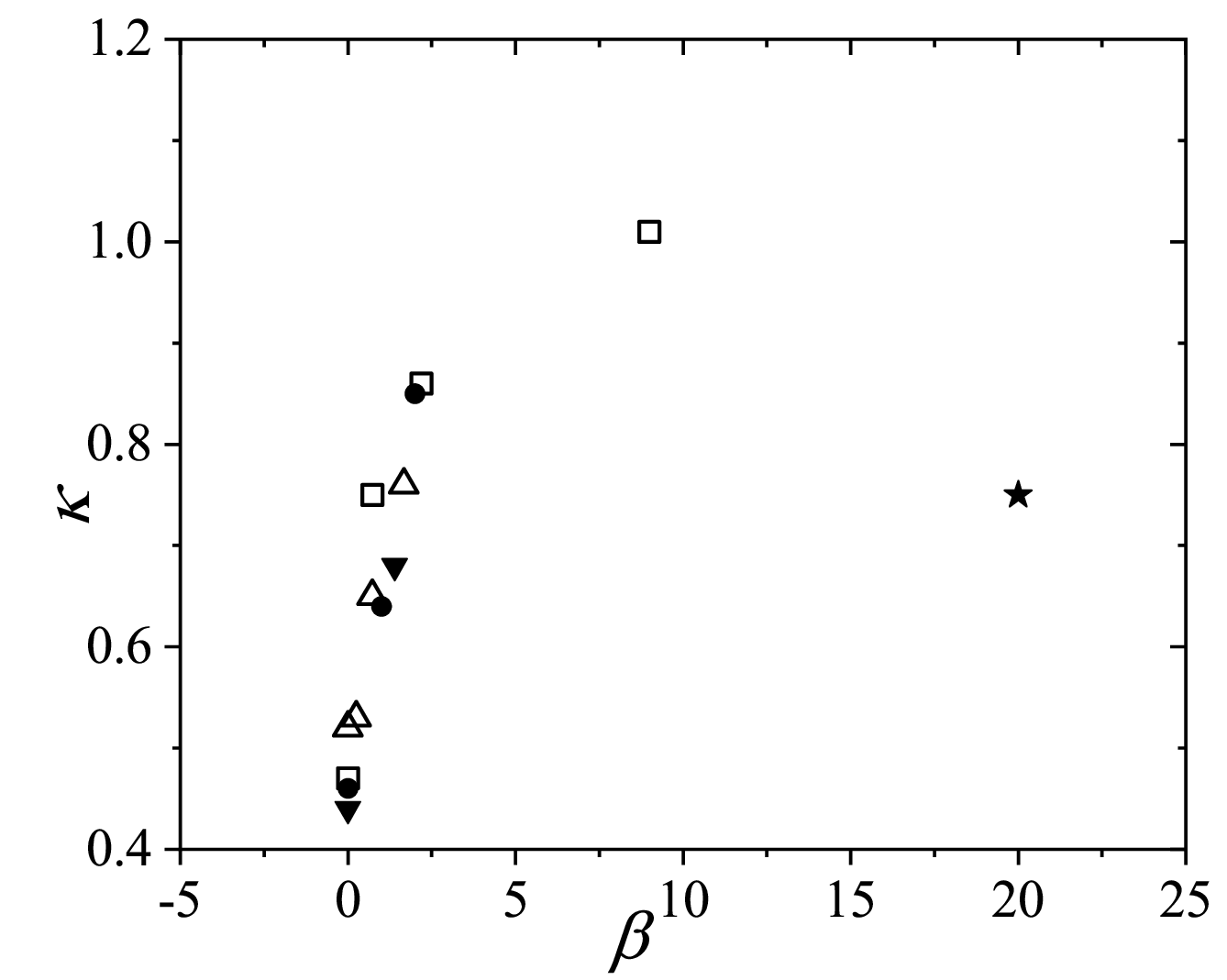}
    \caption{}
  \end{subfigure}
\begin{center}
  \begin{subfigure}[b]{0.5\linewidth}
    \centering
    \includegraphics[width=\linewidth]{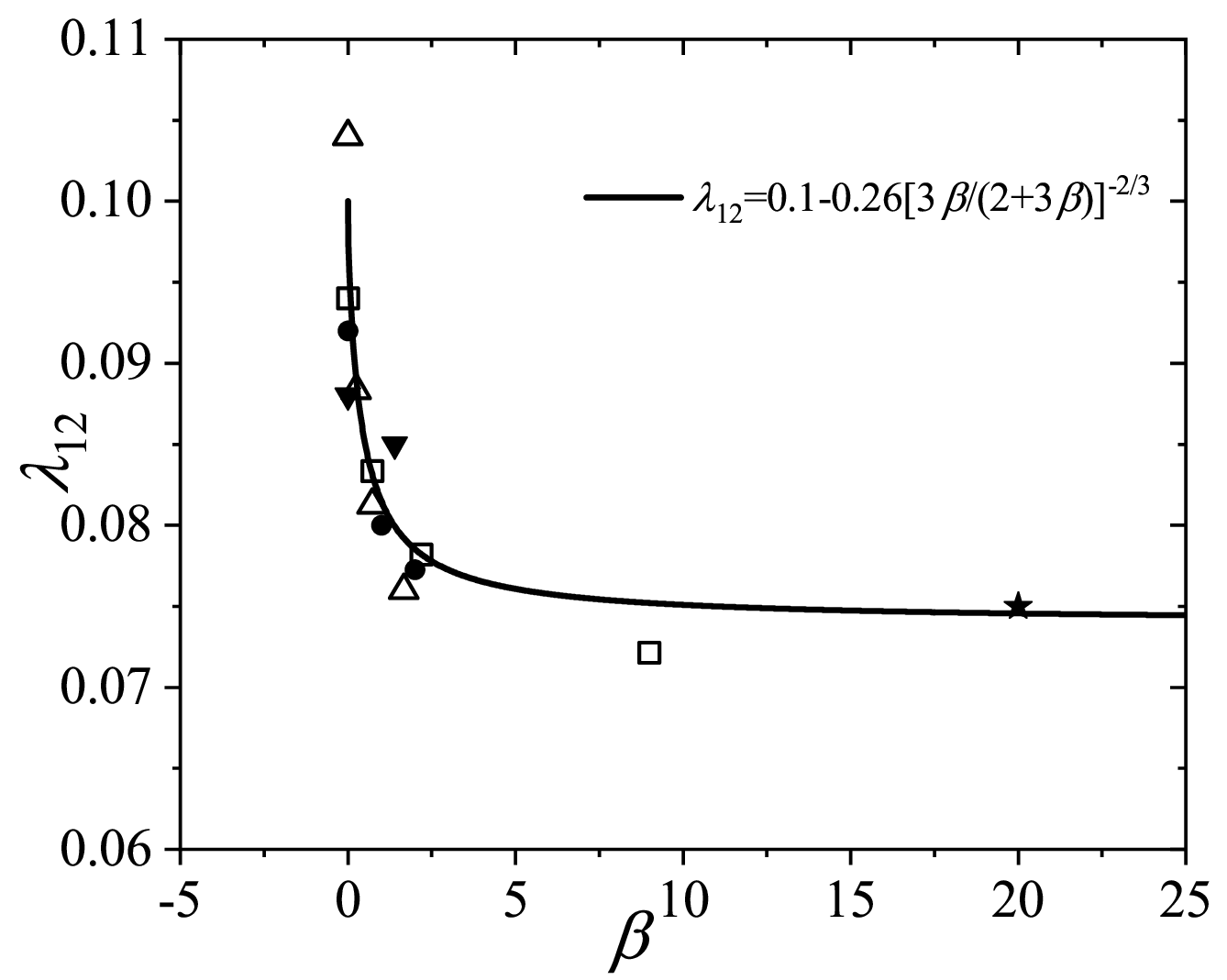}
    \caption{}
  \end{subfigure} 
\end{center}
  \caption{Dependence of (a) $n$, (b) $\kappa_{12}$, and (c) $\lambda_{12}$ on $\beta$ for the databases in table \ref{tab:cases}.}
  \label{fig:n_kappa_yb_beta_Retau}
\end{figure}

The variations of $n$ and $\kappa_{12}$ with $\beta$ are shown in figures \ref{fig:n_kappa_yb_beta_Retau}(a) and \ref{fig:n_kappa_yb_beta_Retau}(b). At small and moderate $\beta$, both $n$ and $\kappa_{12}$ increase rapidly with increasing $\beta$, representing more and more extended constant-stress-length region as well as more and more rapid increase of $\ell_{12}$ with increasing $y$ near the wall (figure \ref{fig:lm_result}). It is interesting that the large-$\beta$, high-$Re_\tau$ case of \cite{Skare1994exp} seems to disengage from the trend of the other equilibrium APG TBLs, even when the $\beta=9$ case of \cite{Lee2017direct} is considered untypical. Therefore, more moderate $\beta$, high Reynolds number data are needed to fill the gap between the case of \cite{Skare1994exp} and the others, in order to obtain a full picture. \cite{Skare1994exp} also observed the rapid increase of $\kappa_{12}$ with increasing $\beta$. They marked the $\kappa_{12}$ in (\ref{eq:length_out}) with $\kappa_\ell$ and the $\kappa$ in (\ref{eq:U_wake}) with $\kappa_u$, and pointed out that, while $\kappa_\ell$ varies dramatically with $\beta$, $\kappa_u$ keeps the same as that in the ZPG TBL. Their observation agrees with ours as shown in figure \ref{fig:U_models} for the $\beta=1.4$ case of \cite{Pozuelo2022les}, in which $\kappa_u=0.41$ and $\kappa_\ell=0.68$. This reveals an interesting doubt that the mixing length hypothesis and the logarithmic law of the wall seem to be less correlated in APG TBLs. 

The similar variations of $n$ and $\kappa_{12}$ with $\beta$ reveals that $\lambda_{12}$ ($=\kappa_{12}/n$) may change less when $\beta$ varies. As shown in figure \ref{fig:n_kappa_yb_beta_Retau}(c), $\lambda_{12}$ decreases with increasing $\beta$, meaning that the relative eddy size becomes smaller when the APG is strengthened. In view of the similarity between figure \ref{fig:n_kappa_yb_beta_Retau}(c) and figures \ref{fig:uvmax_P0_beta}(c) and \ref{fig:uvmax_P0_beta}(d), we can construct an empirical model for $\lambda_{12}$ as below
\begin{equation}
  \lambda_{12}\approx0.1-0.026\left(\frac{3\beta}{2+3\beta}\right)^{2/3}.
  \label{eq:lambda12}
\end{equation}
As shown in figure \ref{fig:n_kappa_yb_beta_Retau}(c), the model agrees with the data rather well. We thus deduce that $\lambda_{12}$ is about 0.074 when $\beta$ is considerably large. This finding reveals that an ultimate state independent of $\beta$ may occur for the stress length and mean velocity profiles in the wake region. 

\begin{figure}
  \centering
    \includegraphics[width=0.5\linewidth]{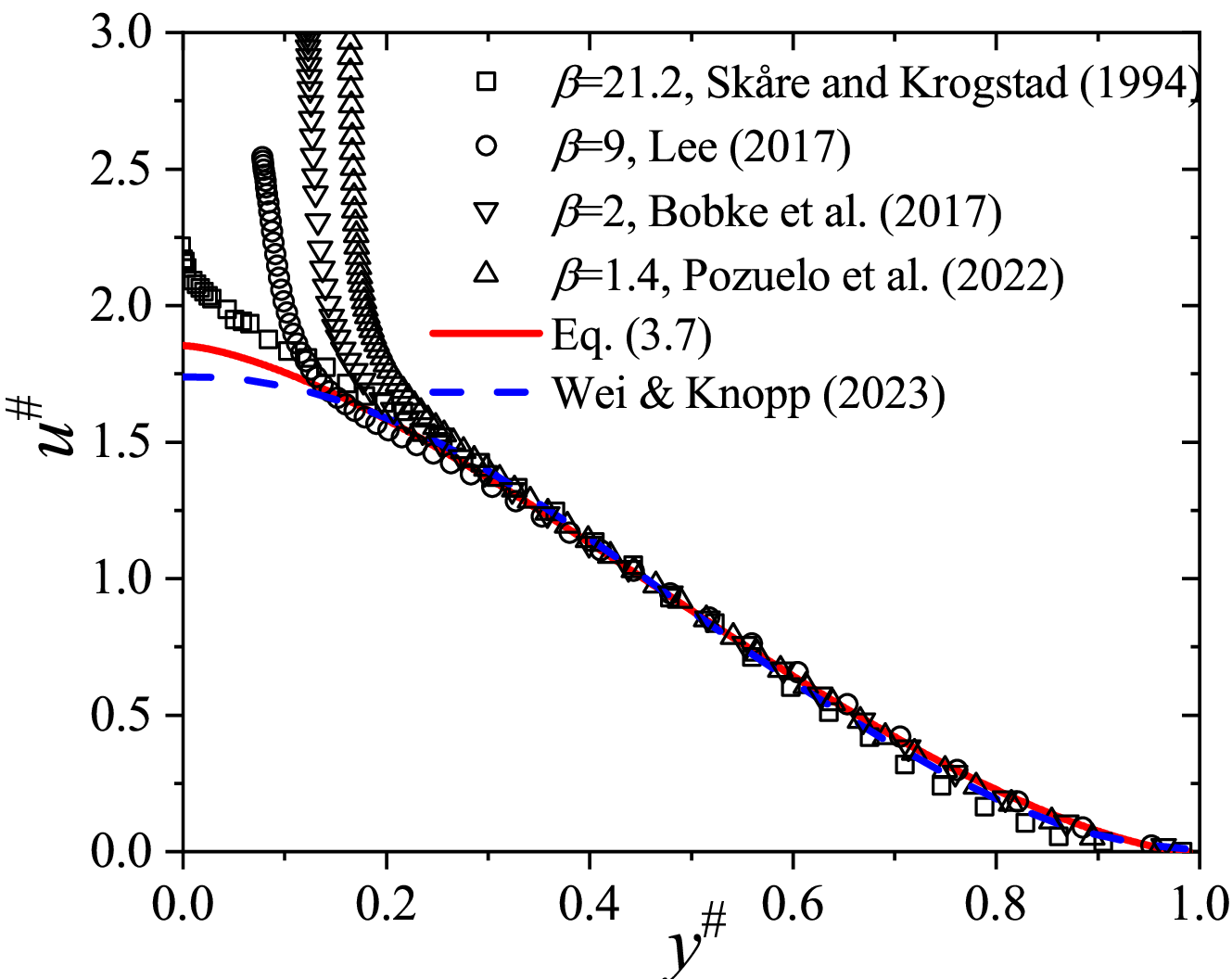}
  \caption{The invariant outer-layer mean velocity profile predicted by (\ref{eq:udefect_outscaling}), compared with the empirical profile of \cite{wei_knopp_2023} and the databases in table \ref{tab:cases}.}
  \label{fig:udefect_outscaling}
\end{figure}

To derive the ultimate-state mean velocity deficit profile, we consider only the constant-stress-length region, which covers a majority portion of the boundary layer as shown in figure \ref{fig:lm_result}. In the constant-stress-length region, according to (\ref{eq:udefect_model}),
 \begin{equation}
  u_\delta^+-u^+=\int_{y^*}^{1}{\frac{\sqrt{\tau^+}}{\lambda_{12}}\mathrm{d} y^*}=\frac{\tau_{\rm max}^+(1-y_{\rm max}^*)}{0.545\lambda_{12}}\int_{y^\#}^{1}{\sqrt{\tau^\#}\mathrm{d} y^\#}.
 \label{eq:udefect_inv}
 \end{equation}
Subsequently, a ultimate-state mean velocity profile can be constructed as
 \begin{equation}
  u^\#=\frac{u_\delta^+-u^+}{u_\delta^+-u_m^+}=\frac{\int_{y^\#}^{1}{\sqrt{\tau^\#}\mathrm{d} y^\#}}{\int_{0.455}^{1}{\sqrt{\tau^\#}\mathrm{d} y^\#}},
 \label{eq:udefect_outscaling}
 \end{equation}
where $u_m^+$ is the mean velocity at $y_{\rm max}^*$. (\ref{eq:udefect_outscaling}) is the outer scaling of the streamwise mean velocity proposed by \cite{wei_knopp_2023} via a scaling patch approach. Through fitting multiple datasets \cite{wei_knopp_2023} obtained an empirical profile that reads 
$(u_\delta-u)/(u_\delta-u_m)=1-erf(1.3y^{**}+0.21(1.3y^{**})^4)$, where $y^{**}=(y^\#-0.455)/0.545$ as aforementioned. Our study presents the analytical expression of $u^\#$. Figure \ref{fig:udefect_outscaling} validates the invariance of $u^\#$ with the databases in table \ref{tab:cases}. In the wake region, (\ref{eq:udefect_outscaling}) agrees with the data and the empirical profile of \cite{wei_knopp_2023} quite well.

The Coles' law of the wake is widely recognized because of its simplicity and reasonable accuracy in describing 2D equilibrium and even nonequilibirum PG TBLs. In the history it stands as a benchmark model for validating measurements and turbulence models. Therefore, we present more analysis, aiming at revealing a consistency between the current theory and the Coles' law of the wake.

\begin{figure}
  \centering
    \includegraphics[width=0.5\linewidth]{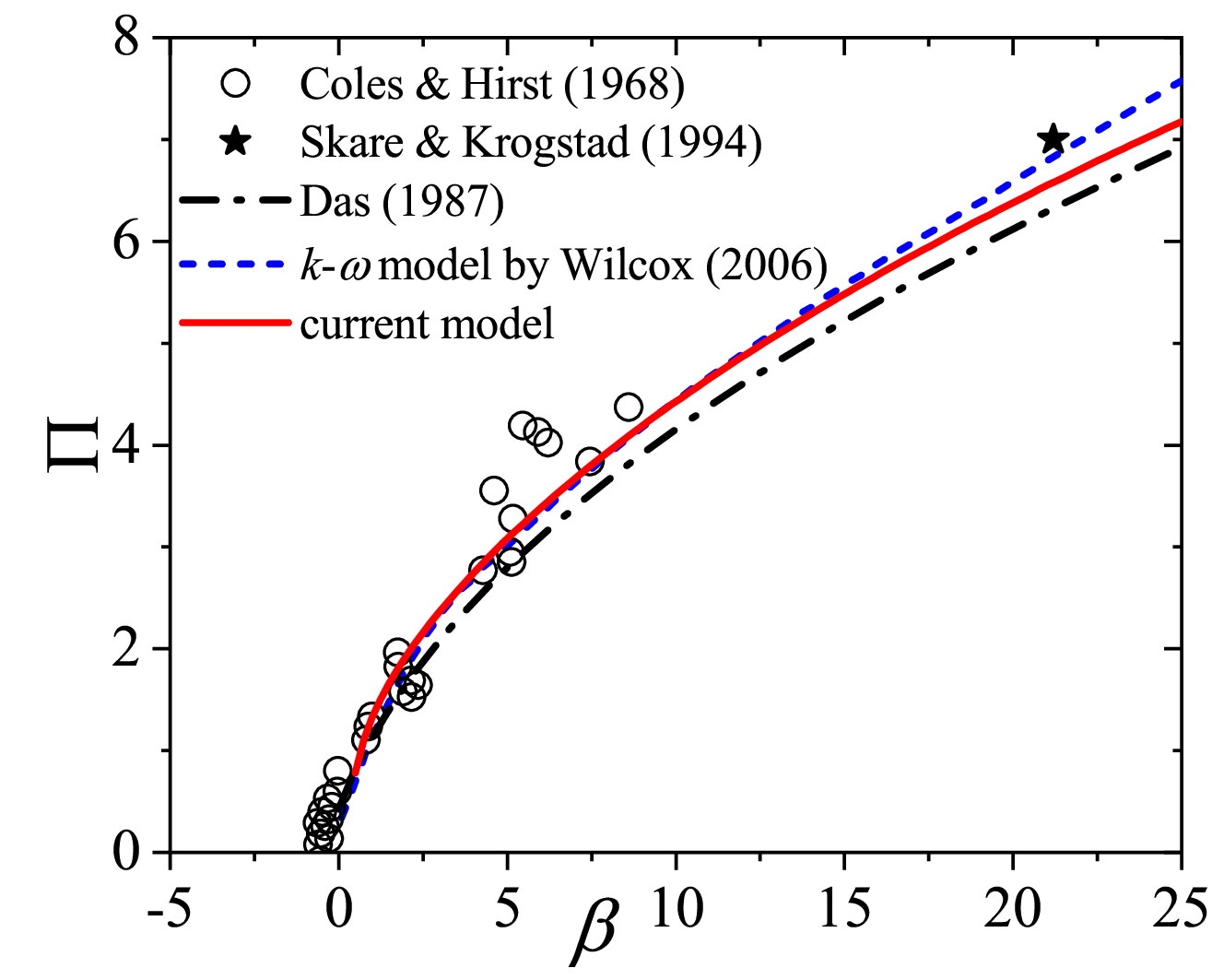}
  \caption{The wake parameter $\Pi$ as a funtion of $\beta$ for different models compared with experiments.}
  \label{fig:Pi_beta}
\end{figure}

According to the Coles' law of the wake, the mean shear in the wake region reads: 
\begin{equation}
  S^+(y^*)=\frac{1}{\kappa y^*}+\frac{\pi\Pi}{\kappa}{\rm sin}(\pi y^*),
  \label{eq:shear_coles}
\end{equation}
Meanwhile, in the wake region (\ref{eq:mean_shear}) $S^+(y^+)=\sqrt{\tau^+}/\ell_{12}^+$ in which $\tau^+$ and $\ell_{12}^+$ are described by (\ref{eq:tau_PG}) and (\ref{eq:length_out}), respectively. Because both the Coles' law of the wake and the current theory rather accurately describe the mean velocity profile (as shown in figure \ref{fig:U_models}), one may expect that the two mean-shear descriptions are quite close to each other in the wake region, even though it is difficult to prove it analytically. To simplify the derivation, we assume that the two mean-shear descriptions are equal at $y_{\rm max}$. Consequently,
\begin{equation}
  \frac{1}{\kappa y_{\rm max}^*}+\frac{\pi\Pi}{\kappa}{\rm sin}(\pi y_{\rm max}^*)=\frac{\sqrt{1+0.75\beta}}{\lambda_{12}}.
  \label{eq:Pi_model}
\end{equation}
where $\kappa=0.41$, $y_{\rm max}^*$ is described by (\ref{eq:ymax}), and $\lambda_{12}$ is described by (\ref{eq:lambda12}). (\ref{eq:Pi_model}) presents a new model for the wake parameter $\Pi(\beta)$. As shown in figure \ref{fig:Pi_beta}, the current model agrees quite well with the experimental data, the empirical model of \cite{Das1987}, and the prediction of the $k-\omega$ turbulence model of \cite{Wilcox2006}. At above moderate $\beta$, $y_{\rm max}^*\approx0.455$ and $\lambda_{12}\approx 0.074$, such that (\ref{eq:Pi_model}) is reduced to $\beta\approx 0.42\Pi^2+0.6\Pi-1.1$, which is very close to the model of \cite{Das1987} which reads $\beta\approx 0.42\Pi^2+0.76\Pi-0.4$. Note that \cite{Das1987} has correlated hundreds of data points from the 1968 Stanford Conference \citep{ColesHirst1968}. Therefore, the current theory is quantitatively consistent with much more experimental data, as well as the celebrated Coles' law of the wake.

\section{Conclusion}\label{sec:conclusion}
The equilibrium PG TBLs have been extensively studied during the past decades. For such benchmark flows with prominent similarity properties, abundant of knowledge has been accumulated with respect to the analytical features, qualitative and quantitative mean-flow behaviors, instantaneous and statistical flow structures, etc. However, a theory that can describe and predict the mean velocity and Reynolds stress components' profiles is still absent, which restricts a deep understanding of the flows as well as development of high-fidelity turbulence models. This study seeks such a theory for the wake region of the equilibrium APG TBL, based on the symmetry approach of the structural ensemble dynamics theory.   

In this paper, we follow the conventional total-stress mixing-length closure strategy to construct the theory. Firstly, we explore the PG-induced dilation-symmetry-breaking of the total stress $\tau^+$ to construct a modified defect power law for $\tau^+$ of the equilibrium APG TBLs. Crucially, a PG stress $P_0^+$ is identified, which quantifies the APG-induced total-stress overshoot and is proportional to the Clauser PG parameter $\beta$. The wall-normal location with peak stress is predicted to be $y_{\rm max}^*=y_{\rm max}/\delta=0.455[3\beta/(a+3\beta)]^{2/3}$. With a transformation of $\tau^\#=\tau^+/\tau_{\rm max}^+$, $y^\#=1-0.545(1-y^*)/(1-y_{\rm max}^*)$, a total stress profile with an arbitrary $\beta$ can be transformed into an invariant profile of $\tau^{\#}=2[{1+{{({0.49/y^\#})}^4}}]^{-1/4}-2y^{\#3/2}$, which is the ultimate state of the total stress at infinite $\beta$. Therefore, we derive an outer scaling for the total stress (and the Reynolds shear stress) of the equilibrium APG TBLs. This outer scaling is equivalent to the recent proposal of \cite{wei_knopp_2023}, but the present invariant profile is proposed theoretically, instead of fitting. Because of the similarity of the Reynolds shear and normal stresses in the wake region, the outer-layer profiles of the streamwise, wall-normal, and spanwise Reynolds normal stresses are predicted accordingly. Especially, the peak stresses in the wake region are captured.  

Secondly, a defect power law is proposed for the stress and kinetic energy length functions in the wake region. Two critical paramters in the defect power law are identified to depend on $\beta$ and determine the profiles of the length functions. One is the exponent $n$, which quantifies the extent of the wake region, and increases rapidly with increasing $\beta$ when $\beta$ is less than about 2.5. The other is $\lambda_{ij}$, which quantifies the relative eddy length in the wake region, and decreases rapidly with increasing $\beta$ when $\beta$ is less than about 2.5, but quickly approaches 0.074 when $\beta$ is larger. The streamwise mean-velocity-deficit profile in the wake region of the equilibrium APG TBL is predicted with the defect power law of the stress length and the total stress model. In addition, an invariant mean velocity profile is derived, which describes the ultimate state of the mean velocity in the wake region at infinite-$\beta$. This invariant mean velocity profile is also equivalent to the proposal of \cite{wei_knopp_2023}, but the present one is proposed theoretically, instead of fitting. Furthermore, with the current theory, the variation of the Coles' wake parameter $\Pi$ with $\beta$ is predicted, and in close agreement with the empirical relation of \cite{Das1987} that correlates hundreds of experimental data.  

The above predictions are validated with five published DNS, LES, and experimental databases on the equilibrium APG TBLs. The theory agrees with the data rather well over wide ranges of $\beta$ and Reynolds number.

The success of the theory reveals that the dilation symmetry is an important property of TBLs, and can be employed for constructing mean-field theory of complex TBLs. Especially, the universal dilation ansatz is crucial for describing the dilation-symmetry-breaking induced by a complicated boundary-layer effect. Owing to the universality of the dilation symmetry and dilation-symmetry-breaking in boundary-layer flows, the theory can also be extended to describe nonequilibrium PG TBLs, as demonstrated by our recent study on separation bubble flow. In subsequent studies, we will discuss the PG effects on the inner flow of the equilibrium PG TBLs, and present how the formulations of total stress and length functions can be extended to capture typical flow-history effects, such that a general PG TBL can be analyzed theoretically.   

The authors are grateful to Prof. J.-H. Lee for sharing his DNS data. This research is supported by the National Natural Science Foundation of China under Grant No. 91952201.

%\backsection[Declaration of interests]{The authors report no conflict of interest.}

\bibliographystyle{jfm}
\bibliography{jfm-main}

\end{document}